\documentclass[aps,prd,preprint,superscriptaddress,
preprintnumbers,eqsecnum,nofootinbib,nobibnotes,showpacs]{revtex4}

\usepackage{amsfonts,amssymb,amsmath}
\usepackage{graphicx}

\newcommand{\be}{\begin{equation}}
\newcommand{\bea}{\begin{eqnarray}}
\newcommand{\ee}{\end{equation}}
\newcommand{\eea}{\end{eqnarray}}
\newcommand{\bpi}{\begin{picture}}
\newcommand{\bce}{\begin{center}}
\newcommand{\epi}{\end{picture}}
\newcommand{\ece}{\end{center}}

\def\chic#1{{\scriptscriptstyle #1}}

\def\NV{\bqq'}

\def\NP{ V}

\def\bqq{{\mathrm{I}\!\Gamma}}
\def\ff{B}

\def\bcj{J}
\def\bcjb{\widetilde{\bcj}}

\begin{document}

\title{Massless bound-state excitations \\  and the Schwinger mechanism in QCD}

\author{A.C.~Aguilar}
\affiliation{Federal University of ABC, CCNH, \\
Rua Santa Ad\'{e}lia 166, CEP 09210-170, Santo Andr\'{e}, Brazil.}

\author{D.~Ib\'a\~nez}
\affiliation{Department of Theoretical Physics and IFIC, 
University of Valencia-CSIC,
E-46100, Valencia, Spain.}

\author{V.~Mathieu}
\affiliation{Department of Theoretical Physics and IFIC, 
University of Valencia-CSIC,
E-46100, Valencia, Spain.}

\author{J.~Papavassiliou}
\affiliation{Department of Theoretical Physics and IFIC, 
University of Valencia-CSIC,
E-46100, Valencia, Spain.}

\begin{abstract}

The gauge invariant generation of an effective gluon mass 
proceeds through the well-known Schwinger mechanism, 
whose key dynamical ingredient is the nonperturbative formation 
of longitudinally coupled massless bound-state excitations.
These excitations introduce poles in the 
vertices of the theory, in such a way as to 
maintain the Slavnov-Taylor identities intact in the presence of 
massive gluon propagators. 
In the present work we first focus on the modifications induced to the 
nonperturbative three-gluon vertex by the inclusion 
of massless two-gluon bound-states into the kernels appearing 
in its skeleton-expansion.  
Certain general relations between the basic building blocks of these bound-states 
and the gluon mass are then obtained 
from the Slavnov-Taylor identities and the 
Schwinger-Dyson equation governing the gluon propagator.
The homogeneous Bethe-Salpeter equation 
determining the wave-function of the aforementioned bound state 
is then derived, under certain simplifying assumptions.
It is then shown, through a detailed analytical and numerical study, 
that this equation admits non-trivial solutions, indicating that 
the QCD dynamics support indeed 
the formation of such massless bound states.
These solutions are subsequently used, in conjunction with the aforementioned relations,  
to determine the momentum-dependence of the dynamical gluon mass.  
Finally, further possibilities and open questions are briefly discussed.

\end{abstract}


\maketitle
\section{\label{Int}Introduction}

The numerous large-volume lattice simulations carried out in recent years 
have firmly established that, in the Landau gauge,  
the  gluon propagator and  the  ghost   dressing  function  of  pure  Yang-Mills
theories are  infrared 
finite,    both     in    
$SU(2)$~\cite{Cucchieri:2007md,Cucchieri:2007rg,Cucchieri:2009zt,Cucchieri:2011ga,Cucchieri:2011um}
and    in
$SU(3)$~\cite{Bogolubsky:2007ud,Bowman:2007du,Bogolubsky:2009dc,Oliveira:2009eh}.
Perhaps the most physical way of explaining the observed 
finiteness of these  quantities is 
the generation of a non-perturbative, momentum-dependent 
gluon mass~\cite{Cornwall:1981zr,Aguilar:2002tc,Aguilar:2006gr,Aguilar:2008xm,Aguilar:2009ke,Aguilar:2011ux}, 
which acts as a natural infrared cutoff.   
In this picture  
the fundamental Lagrangian of the Yang-Mills theory (or that of QCD) remains unaltered, and   
the generation of the gluon mass takes place dynamically, 
through the well-known Schwinger mechanism~\cite{Schwinger:1962tn,Schwinger:1962tp,Jackiw:1973tr,Jackiw:1973ha,Cornwall:1973ts,Eichten:1974et,Poggio:1974qs} 
without violating any of the underlying symmetries
(for further studies and alternative approaches, see, {\it e.g.},
\cite{Boucaud:2008ji,Dudal:2008sp,Fischer:2008uz,Kondo:2009gc,Oliveira:2010xc,RodriguezQuintero:2010ss,Pennington:2011xs}).

The way how 
the Schwinger mechanism generates a  mass for the gauge boson (gluon)  
can be seen most directly at the level of its 
inverse propagator, 
$\Delta^{-1}({q^2})=q^2 [1 + i {\bf \Pi}(q^2)]$, where ${\bf \Pi}(q)$ 
is the dimensionless vacuum polarization.
According to Schwinger's fundamental observation, 
if ${\bf \Pi}(q^2)$ 
develops a pole at zero momentum transfer ($q^2=0$), then the 
vector meson acquires a mass, even if the gauge symmetry 
forbids a mass term at the level of the fundamental Lagrangian.
Indeed, if ${\bf \Pi}(q^2) = m^2/q^2$, then (in Euclidean space)
\mbox{$\Delta^{-1}(q^2) = q^2 + m^2$}, and so 
the vector meson becomes massive, $\Delta^{-1}(0) = m^2$, 
even though it is massless in the absence of interactions 
($g=0$, ${\bf \Pi} =0$)~\cite{Jackiw:1973tr,Jackiw:1973ha}.

The key assumption when invoking the Schwinger mechanism in Yang-Mills theories, 
such as QCD, is that the required  poles   may be produced 
due to purely dynamical reasons; specifically, one assumes that, for sufficiently 
strong binding, 
the mass of the appropriate bound state may be reduced to zero~\cite{Jackiw:1973tr,Jackiw:1973ha,Cornwall:1973ts,Eichten:1974et,Poggio:1974qs}. 
In addition to triggering the Schwinger mechanism, 
these massless composite excitations are crucial for preserving gauge invariance. 
Specifically, the presence of massless poles  in the off-shell interaction vertices 
guarantees that the Ward identities (WIs) and Slavnov Taylor identities (STIs) of the theory 
maintain exactly the same form before and after mass generation (i.e. when the 
the massless propagators appearing in them are replaced by massive ones)
~\cite{Cornwall:1981zr,Eichten:1974et,Poggio:1974qs,Aguilar:2011ux}.
Thus, these excitations act like dynamical Nambu-Goldstone scalars,
displaying, in fact, all their typical characteristics, such as 
masslessness, compositeness, and longitudinal coupling; 
note, however, that they differ from Nambu-Goldstone bosons 
as far as their origin is concerned, since they are not associated 
with the spontaneous breaking of any global symmetry~\cite{Cornwall:1981zr}.
Finally, every such Goldstone-like scalar, ``absorbed'' by
a gluon in order to acquire a mass,  
is expected to actually cancel out of the $S$-matrix   
against other massless poles or due to current conservation~\cite{Jackiw:1973tr,Jackiw:1973ha,Cornwall:1973ts,Eichten:1974et,Poggio:1974qs}.

The main purpose of the present article is to scrutinize the central assumption of  
the dynamical scenario outlined above, namely the possibility of actual formation 
of such massless excitations. The question we want to address is 
whether the non-perturbative Yang Mills dynamics 
are  indeed compatible with the generation of such a special bound-state.  
In particular, as has already been explained in previous works, 
the entire mechanism of gluon mass generation hinges on the 
appearance of massless poles inside the nonperturbative three-gluon vertex, 
which enters in the Schwinger Dyson equation (SDE) governing the gluon propagator. 
These poles correspond to the propagator of the scalar massless excitation, 
and interact with a pair of gluons through a very characteristic proper vertex, 
which, of course, must be non vanishing, or else the entire construction collapses.
The way to establish the existence of this latter vertex  is 
through the study of the homogeneous Bethe-Salpeter equation (BSE) that it satisfies, 
and look for  non-trivial solutions, subject to the numerous stringent constraints imposed by 
gauge invariance.

This particular methodology has been adopted in various early contributions on this 
subject; however, only asymptotic solutions to the corresponding equations have been considered.
The detailed numerical study 
presented here demonstrates that, under certain simplifying assumptions 
for the structure of its kernel,  
the aforementioned integral equation 
has indeed non-trivial solutions, valid for the entire range of physical momenta. 
This result, although approximate and not fully conclusive, furnishes additional 
support in favor of the concrete mass generation mechanism described earlier.

The article is organized as follows.
In Section \ref{gc} we set up the general theoretical framework related to the 
gauge-invariant generation of a gluon mass; in particular, 
we outline how the vertices of the theory must be modified, through the inclusion 
of longitudinally coupled massless poles,  
in order to maintain the  WIs and STIs of the theory intact.
In Section \ref{pv} we take a detailed look into the structure of the 
non-perturbative vertex that contains the required massless poles, and study its main 
dynamical building blocks, and in particular the transition amplitude between 
a gluon and a massless excitation 
and the proper vertex function (bound-state wave function), controlling the interaction 
of the massless excitation with two gluons. 
In addition, we derive an exact relation between these two 
quantities and the first derivative of the (momentum-dependent) gluon mass.  
Then, we derive a simple formula that, at zero momentum transfer,  relates 
the aforementioned transition amplitude to the  gluon mass.
In the next two sections we turn to the central question  of this work, namely the 
dynamical realization of the massless excitation within the Yang-Mills theory.
Specifically, in Section \ref{bse} 
we derive the BSE that the proper vertex function satisfies,  
and implement a number of simplifying assumptions. Then,  
in Section \ref{ns} we demonstrate through a detailed numerical study 
that the resulting homogeneous integral equation 
admits indeed non-trivial solutions, thus corroborating the existence of the 
required bound-state excitations. 
In Section \ref{dec} we demonstrate with a specific example
the general mechanism that leads to the decoupling of all massless poles  
from the physical (on-shell) amplitude.
Finally, in Section \ref{conc} we discuss our results and present our conclusions.

\section{\label{gc}General considerations}

In this section, after establishing the necessary notation, we briefly review 
why the dynamical generation of a mass is inextricably connected to the presence of a 
special vertex, which exactly compensates for the appearance of massive instead 
of massless propagators in the corresponding  WIs and STIs.

The full gluon propagator 
$\Delta^{ab}_{\mu\nu}(q)=\delta^{ab}\Delta_{\mu\nu}(q)$ in the Landau gauge is defined as
\be
\Delta_{\mu\nu}(q)=- i P_{\mu\nu}(q)\Delta(q^2) \,,
\label{prop}
\ee 
where
\be
P_{\mu\nu}(q)=g_{\mu\nu}- \frac{q_\mu q_\nu}{q^2} \,,
\ee
is the usual transverse projector, 
and the scalar cofactor $\Delta(q^2)$  
is related to the (all-order) gluon self-energy $\Pi_{\mu\nu}(q)=P_{\mu\nu}(q)\Pi(q^2)$  through
\be
\Delta^{-1}({q^2})=q^2+i\Pi(q^2).
\label{defPi}
\ee
One may define the dimensionless vacuum polarization ${\bf \Pi}(q^2)$ 
by setting  $\Pi(q^2) = q^2 {\bf \Pi}(q^2)$ so that (\ref{defPi}) becomes 
\be
\Delta^{-1}({q^2})=q^2 [1 + i {\bf \Pi}(q^2)] \,.
\label{defvp}
\ee
As explained in the Introduction, if ${\bf \Pi}(q^2)$ develops  
at zero momentum transfer a pole with positive residue $m^2$, then 
$\Delta^{-1}({0}) = m^2$, and the 
gluon is endowed with an  effective mass. 

Alternatively, one may define the gluon dressing function $J(q^2)$ as 
\be
\Delta^{-1}({q^2})=q^2 J(q^2) \,.
\label{defJ}
\ee
In the presence of a dynamically generated mass, the natural form of $\Delta^{-1}(q^2)$ is given by 
(Euclidean space) 
\be
\Delta^{-1}(q^2) =q^2 J(q^2) + m^2(q^2) \,,
\label{defm}
\ee
where the first term corresponds to the ``kinetic term'', or ``wave function'' contribution, 
whereas the second is the (positive-definite) momentum-dependent mass.
If one insist on maintaining the form of (\ref{defJ}) by explicitly factoring out a $q^2$, then   
\be
\Delta^{-1}({q^2})=q^2  \left[J(q^2) + \frac{m^2(q^2)}{q^2}\right]\,,
\label{defJm}
\ee
and the presence of the pole, with residue given by $m^2(0)$, becomes manifest.


\begin{figure}[!t]
\includegraphics[scale=0.8]{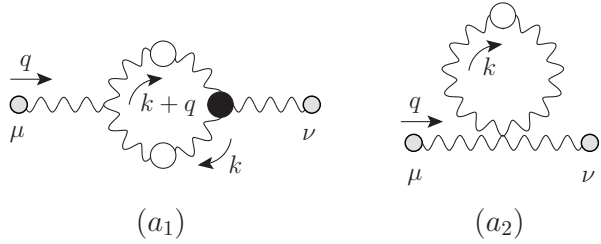}
\caption{\label{gSDE} The ``one-loop dressed'' gluon contribution to the  PT-BFM gluon self-energy. 
White (black) circles denote fully dressed propagators (vertices); 
a gray circle attached to the external legs indicates that they are background gluons. 
Within the PT-BFM framework these 
two diagrams constitute a transverse subset of the full gluon SDE.}
\end{figure}


Of course, in order to obtain the full dynamics, 
such as, for example, the momentum-dependence of the dynamical mass, 
one must turn eventually  to the SDE that governs 
the corresponding gauge-boson self-energy (see Fig.~\ref{gSDE}). 
In what follows we will work within 
the specific framework provided by the  synthesis of the pinch technique (PT)
~\cite{Cornwall:1981zr,Cornwall:1989gv,Pilaftsis:1996fh,Binosi:2002ft,Binosi:2003rr,Binosi:2009qm}
with the background field method (BFM)~\cite{Abbott:1980hw}. 
One of the main advantages of the ``PT-BFM'' formalism is that 
the crucial transversality property 
of the gluon self-energy $\Pi_{\mu\nu}(q)$, namely $q^{\mu}\Pi_{\mu\nu}(q)=0$ ,
is maintained at the level of the {\it truncated} SDEs~\cite{Aguilar:2006gr,Binosi:2008qk}.

The Schwinger mechanism
is integrated into the SDE of the gluon propagator through the form of the three-gluon vertex.
In particular, as has been emphasized in some of the literature cited above (e.g.,\cite{Aguilar:2011ux}),  
a crucial condition for the realization of the gluon mass generation scenario  
is the existence of a special vertex, to be denoted by $V_{\alpha\mu\nu}(q,r,p)$  
which must be completely {\it longitudinally coupled}, 
i.e. must satisfy 
\be
P^{\alpha'\alpha}(q) P^{\mu'\mu}(r) P^{\nu'\nu}(p) V_{\alpha\mu\nu}(q,r,p)  = 0 \,.
\label{totlon}
\ee 
We will refer to this special vertex 
as the ``pole vertex'' or simply ``the vertex $V$''. 

The role of the vertex $V_{\alpha\mu\nu}(q,r,p)$ is indispensable for maintaining gauge invariance, 
given that 
the massless poles that it must contain in order to trigger the Schwinger mechanism, 
act, at the same time, as composite, longitudinally coupled Nambu-Goldstone bosons. 
Specifically, in order to preserve the gauge-invariance of the theory in the presence of masses, 
the vertex $V_{\alpha\mu\nu}(q,r,p)$ must be added to the 
conventional (fully-dressed) three-gluon vertex $\bqq_{\alpha\mu\nu}(q,r,p)$, giving rise 
to the new full vertex,  $\NV_{\alpha\mu\nu}(q,r,p)$, defined as  
\be
\NV_{\alpha\mu\nu}(q,r,p) = \bqq_{\alpha\mu\nu}(q,r,p) +V_{\alpha\mu\nu}(q,r,p)\,.
\label{NV}
\ee
Gauge-invariance remains intact because $\NV$
satisfies the same STIs as $\bqq$ before, but now replacing the gluon propagators appearing on their
rhs by a massive ones; schematically, $\Delta^{-1} \to \Delta_m^{-1}$, where the former denotes  
the propagator given in (\ref{defJ}), while the latter that of  (\ref{defm}). 
In particular, in the PT-BFM framework that we  employ, 
the vertex $\bqq$ connects a background gluon ($B$) with two quantum gluons ($Q$), and 
is often referred to as the ``BQQ'' vertex. This vertex 
satisfies a (ghost-free) WI when contracted with the momentum $q_\alpha$ of the 
background gluon, whereas it satisfies a STI when contracted with 
the momentum $r_\mu$ or $p_\nu$ of the quantum gluons. In particular, 
\bea
q^\alpha\bqq_{\alpha\mu\nu}(q,r,p)&=&p^2\bcj(p^2)P_{\mu\nu}(p)-r^2\bcj(r^2)P_{\mu\nu}(r),
\nonumber \\
r^\mu\bqq_{\alpha \mu \nu}(q,r,p)&=&F(r^2)\left[q^2\bcjb(q^2)P_\alpha^\mu(q)H_{\mu\nu}(q,r,p)-
p^2\bcj(p^2)P_\nu^\mu(p)\widetilde{H}_{\mu\alpha}(p,r,q)\right], \nonumber \\
p^\nu\bqq_{\alpha \mu \nu}(q,r,p)&=&F(p^2)\left[r^2\bcj(r^2)P_\mu^\nu(r)\widetilde{H}_{\nu\alpha}(r,p,q)-
q^2\bcjb(q^2) P_\alpha^\nu(q)H_{\nu\mu}(q,p,r)\right],
\label{STIs}
\eea
where $F(q^2)$ is the ``ghost dressing function'', defined as  $F(q^2) = q^2 D(q^2)$, 
$H_{\nu\sigma}$ is the standard gluon-ghost kernel, and 
$\widetilde{H}$ is the same as ${H}$ but with   
the external quantum gluon replaced by a background gluon.
Similarly, $\bcjb$ is the dressing function of the self-energy connecting a background with a quantum gluon; 
$\bcjb$ is related to $\bcj(q^2)$ through the identity~\cite{Grassi:1999tp, Binosi:2002ez}
\be
\bcjb(q^2)=\left[1+G(q^2)\right]\bcj(q^2)\,.
\label{BQI}
\ee
The function $G(q^2)$ is  
the scalar co-factor of the $g_{\mu\nu}$ component of the special two-point function 
$\Lambda_{\mu\nu}(q)$, defined as 
\bea
\Lambda_{\mu\nu}(q)&=&-ig^2C_A\int_k\!\Delta_\mu^\sigma(k)D(q-k)H_{\nu\sigma}(-q,q-k,k)\nonumber\\
&=&g_{\mu\nu}G(q^2)+\frac{q_\mu q_\nu}{q^2}L(q^2).
\label{Lambda}
\eea
Note finally that, in the Landau gauge, $G(q^2)$ and  $L(q^2)$ are linked to $F(q^2)$
by the  exact (all-order) relation~\cite{Kugo:1979gm,Grassi:2004yq,Kondo:2009ug,Aguilar:2009nf}
\be
F^{-1}(q^2) = 1+G(q^2) + L(q^2) \,,
\label{funrel}
\ee
to be employed in Subsection D. 

Returning to the nonperturbative vertex $V$,   
gauge invariance requires that 
it must satisfy the WI and STI of (\ref{STIs}), with the replacement $k^2 J(k) \to - m^2(k)$, {\it e.g.},
\be
q^\alpha \NP_{\alpha\mu\nu}(q,r,p)= m^2(r^2)P_{\mu\nu}(r) - m^2(p^2)P_{\mu\nu}(p) \,;
\label{winp}
\ee
exactly analogous 
expressions will hold for the  STIs satisfied when contracting with the  momenta $r$ or $p$.
Indeed, under this assumption, 
the full vertex $\NV$ will satisfy the same WI and STIs
as the vertex $\bqq$ before the introduction of any masses, 
but now with the replacement $ q^2 J(q^2) \to q^2 J(q^2) + m^2(q^2)$. 
Specifically, combining the first relation in (\ref{STIs}) with (\ref{winp}), 
one obtains for the WI of $\NV$, 
\bea
q^{\alpha}\NV_{\alpha\mu\nu}(q,r,p) &=& 
q^{\alpha}\left[\bqq(q,r,p) + \NP(q,r,p)\right]_{\alpha\mu\nu}
\nonumber\\
&=& [p^2\bcj (p^2) -m^2(p^2)]P_{\mu\nu}(p) - [r^2\bcj (r^2) -m^2(r^2)]P_{\mu\nu}(r)
\nonumber\\
&=& \Delta^{-1}_m({p^2})P_{\mu\nu}(p) - \Delta^{-1}_m({r^2})P_{\mu\nu}(r) \,,
\label{winpfull}
\eea
which is indeed the first identity in Eq.~(\ref{STIs}), with the aforementioned replacement 
\mbox{$\Delta^{-1} \to \Delta_m^{-1}$}
enforced. 
The remaining two STIs are realized in exactly the same fashion. 
 
It must be clear at this point that the longitudinal nature of $\NP_{\alpha\mu\nu}$, combined with 
the WI and STIs that it must satisfy, 
lead inevitably to the appearance of a massless pole, as required by the Schwinger mechanism.
For example,  
focusing only on the $q$-channel, the simplest toy Ansatz for the vertex is   
\be
\NP_{\alpha\mu\nu}(q,r,p) = \frac{q_{\mu}}{q^2}[ m^2(r^2)P_{\mu\nu}(r) - m^2(p^2)P_{\mu\nu}(p)]\,,
\ee
which has a pole in $q^2$ and satisfies (\ref{winp}). 
Of course,  poles associated to the other channels ($r$ and $p$) will also appear, given that $\NP_{\alpha\mu\nu}(q,r,p)$ 
must also satisfy the corresponding STIs with respect to $r^\mu$ and $p^\nu$.

\section{\label{pv}The pole vertex: structure and properties}

In this section we have a detailed look at the structure of the special vertex $V$.
In particular, we  
identify the diagrammatic origin and field-theoretic nature of the 
various quantities contributing to it, and specify  
the way it enters into the SDE of the full vertex $\NV$, defined in Eq.~(\ref{NV}). 
In addition, we will derive an exact relation between the most important  
component of this vertex 
and the derivative of the momentum-dependent gluon mass.

\subsection{General structure of the vertex $V$}

The main characteristic of the vertex $V$, 
which sharply differentiates it from ordinary vertex contributions, 
is that it contains massless poles, originating from the 
contributions of bound-state excitations. 
Specifically, all terms of the vertex $V$ are proportional to $1/q^2$, $1/r^2$,  $1/p^2$, 
and products thereof. 
Such dynamically generated poles are to be 
clearly distinguished from poles related to  
ordinary massless propagators, associated with elementary fields in the original Lagrangian.
 
\begin{figure}[ht]
\center{\includegraphics[scale=0.5]{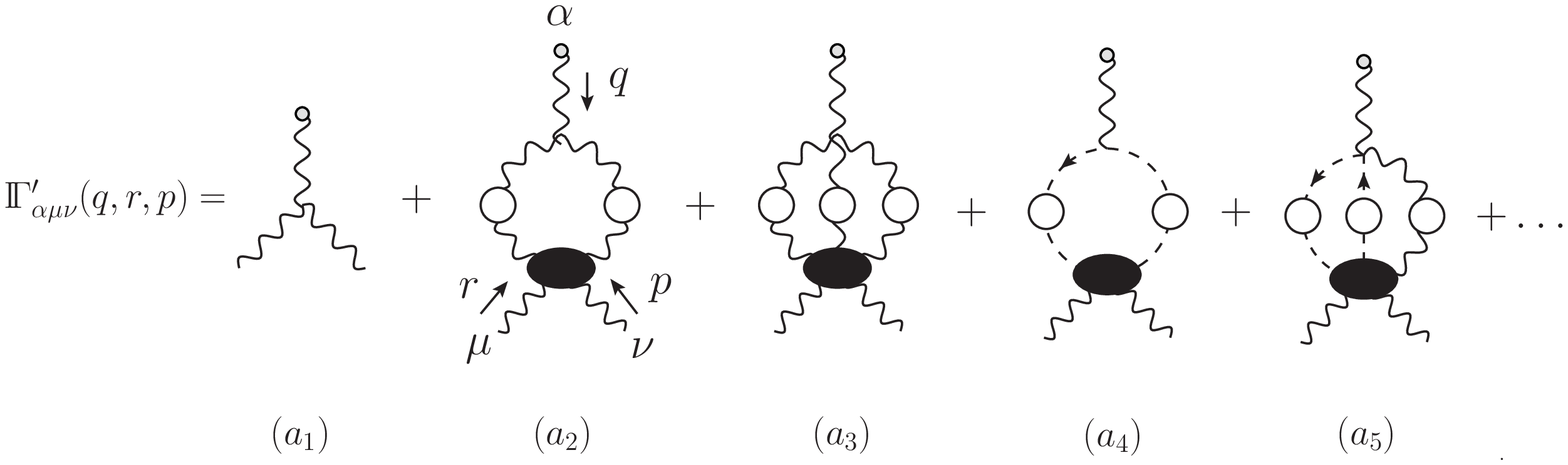}}
\caption{The SDE for the BQQ vertex which connects a background gluon ($B$) with two quantum gluons ($Q$).}
\label{FullvertexSDE}
\end{figure}

\begin{figure}[!t]
\center{\includegraphics[scale=0.5]{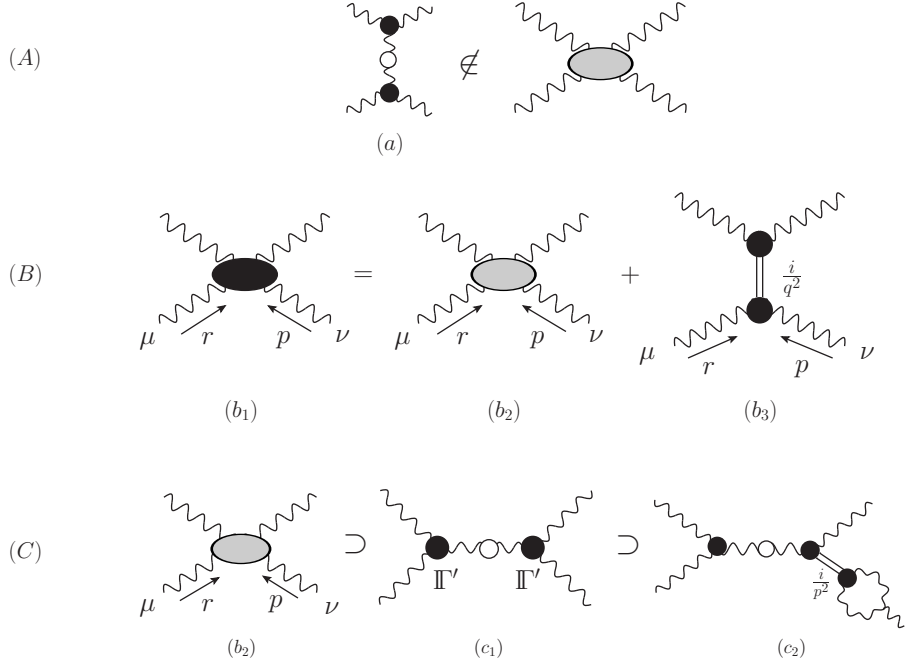}}
\caption{ {\bf(A)} A diagram that does not belong to the standard kernel.
{\bf(B)} The gray kernel (regular part with respect to $q$, and the 
composite massless excitation in the $q$-channel.
{\bf(C)} The $R$ part of the vertex.}
\label{Fullkernel}
\end{figure}

In general, when setting up the usual SDE for any vertex (see, for example, Fig.~\ref{FullvertexSDE}), 
a particular field (leg) is singled out, 
and is connected to the various multiparticle kernels through all 
elementary vertices of the theory involving this field (leg). The remaining legs enter 
into the various diagrams through the aforementioned multiparticle kernels 
(black circles in graphs $a_2$--$a_5$ in Fig.~\ref{FullvertexSDE}),
or, in terms of the standard 
skeleton expansion, through fully-dressed vertices (instead of tree-level ones).  
For the case of the $BQQ$ vertex that we consider here [shown in Fig.~\ref{FullvertexSDE}],  
it is convenient (but not obligatory) to identify 
as the special leg the background gluon, 
carrying momentum $q$. 
Now, with the Schwinger mechanism turned off, the various 
multiparticle kernels appearing in the SDE for the $BQQ$ 
vertex have a complicated skeleton expansion 
(not shown here), 
but their common characteristic is that they are one-particle-irreducible with respect to 
cuts in the direction of the momentum $q$; thus, 
a diagram such as the $(a)$ of Fig.~\ref{Fullkernel} 
is explicitly excluded from the (gray) four-gluon kernel, 
and the same is true for all other kernels.

\begin{figure}[!t]
\center{\includegraphics[scale=0.45]{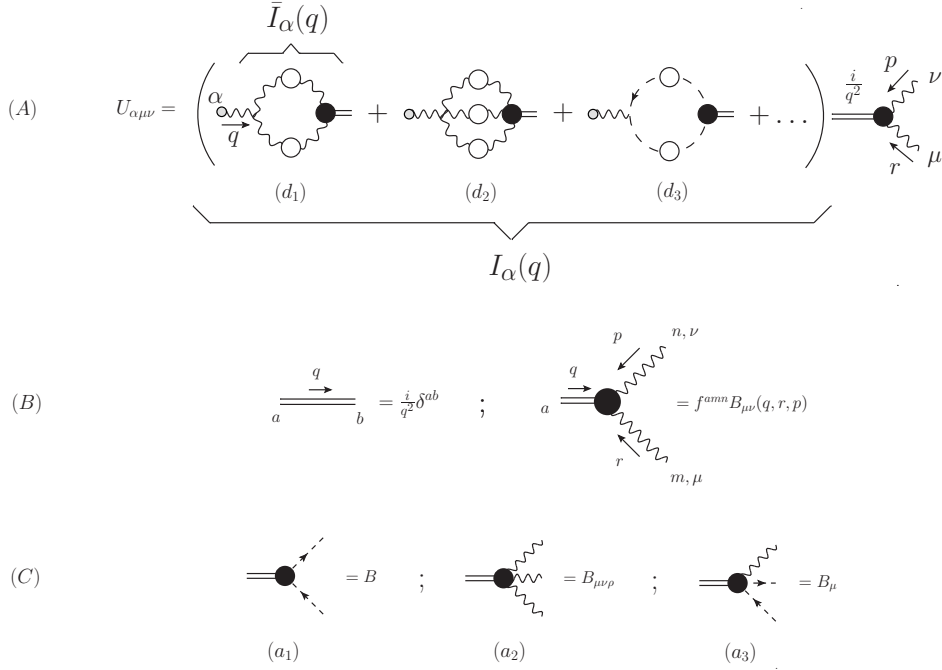}}
\caption{{\bf(A)} The vertex $U$ is composed of three main ingredients: 
the transition amplitude, $I_{\alpha}$, which 
mixes the gluon with a massless excitation, 
the propagator of the  massless excitation, and the (massless excitation)--(gluon)--(gluon) vertex. 
{(\bf B)} The Feynman rules (with color factors included) for (i) the propagator of the massless excitation 
and (ii) the ``proper vertex function'', or, ``bound-state wave function'', $B_{\mu\nu}$.
{(\bf C)} The various $B_{\{\dots\}}$ appearing in Eq.~(\ref{restU}).}  
\label{Uexpansion}
\end{figure}

When the Schwinger mechanism is turned on, 
the structure of the kernels is modified by the presence of composite  
massless excitation, described by a propagator of the type $i/q^2$, as shown in Fig.~\ref{Fullkernel}.
The sum of such dynamical terms, coming from all multiparticle kernels, 
shown in Fig.~\ref{Uexpansion} constitutes a characteristic part of the vertex $V$, 
to be denoted by $U$ in  Eq.~(\ref{VU}),   
namely the part that contains at least a massless propagator $i/q^2$. 
The remaining parts of the vertex $V$, to be denoted by $R$   in  Eq.~(\ref{VR}),
contain massless excitations in the other two channels, 
namely $1/r^2$ and $1/p^2$ (but no $1/q^2$), and originate  
from graphs such as ($c_2$) of Fig.~\ref{Fullkernel}.
Indeed,  note that the kernel ($b_2$) is composed by an infinite number of diagrams, such as ($c_1$),  
containing the full vertex $\NV$; these graphs, 
in turn, will furnish terms proportional to $1/r^2$ and $1/p^2$ [e.g., graph  ($c_2$)].

In order to study further the structure and properties of  the vertex $V$, 
let us first define the full vertex ${\cal V}_{\alpha\mu\nu}^{amn} (q,r,p)$, given by   
\be
{\cal V}_{\alpha\mu\nu}^{amn} (q,r,p) = g f^{amn}  V_{\alpha\mu\nu}(q,r,p) \,, 
\ee
with $V_{\alpha\mu\nu}(q,r,p)$  satisfying Eq.~(\ref{totlon}).
Using a general Lorentz basis, we have the following expansion for $V_{\alpha\mu\nu}(q,r,p)$ 
in terms of scalar form factors, 
\bea 
V_{\alpha\mu\nu}(q,r,p) &=&  
{V}_1 q_{\alpha} g_{\mu\nu}
+  {V}_2 q_{\alpha}q_{\mu}q_{\nu}
+ {V}_3 q_{\alpha} p_{\mu}p_{\nu}
+ {V}_4 q_{\alpha} r_{\mu}q_{\nu}
+ {V}_5 q_{\alpha} r_{\mu}p_{\nu}
\nonumber \\
&+& {V}_6 r_{\mu}g_{\alpha\nu}
+  {V}_7 r_{\alpha}r_{\mu}r_{\nu}
+ {V}_8 r_{\alpha} r_{\mu}p_{\nu}
+ {V}_9  p_{\nu}g_{\alpha\mu}
+ {V}_{10} p_{\alpha}p_{\mu}p_{\nu}\,.
\label{gendec}
\eea

According to the arguments presented above, $V_{\alpha\mu\nu}(q,r,p)$ may be decomposed into 
\be
V_{\alpha\mu\nu}(q,r,p) =  U_{\alpha\mu\nu}(q,r,p) + R_{\alpha\mu\nu}(q,r,p)  \,,
\label{VRU}
\ee
with 
\be
U_{\alpha\mu\nu}(q,r,p)= q_\alpha \bigg(V_1g_{\mu\nu}+V_2q_{\mu}q_{\nu}+V_3p_{\mu}p_{\nu}+
V_4r_{\mu}q_{\nu}+V_5r_{\mu}p_{\nu}\bigg)\quad.
\label{VU}
\ee
and
\be
R_{\alpha\mu\nu}(q,r,p) = \left(V_6g_{\alpha\nu}+V_7r_\alpha r_\nu +\frac{V_8}{2}\, r_\alpha p_\nu\right)r_\mu
+ \left(\frac{V_8}{2}\, r_\alpha r_\mu+V_9g_{\alpha\mu}+V_{10}p_\alpha p_\mu\right)p_\nu \,.
\label{VR}
\ee
All form-factors of $U$ (namely $V_1$--$V_5$) must contain a pole $1/q^2$, while some of them 
may contain, in addition, $1/r^2$ and $1/p^2$ poles.  
On the other hand, none of the form-factors of $R$ (namely $V_6$--$V_{10}$) contains $1/q^2$ poles, but only 
$1/r^2$ and $1/p^2$ poles.

In what follows we will focus on $U_{\alpha\mu\nu}(q,r,p)$, which contains the explicit $q$-channel massless excitation, since 
this is the relevant channel in the SDE of the gluon propagator, 
where  $V_{\alpha\mu\nu}(q,r,p)$ will be eventually inserted [graph $(a_1)$ in Fig.~\ref{gSDE}]. 
In fact, with the two internal gluons of diagram ($a_1$) in the Landau gauge,  we have that 
\bea 
P^{\mu'\mu}(r) P^{\nu'\nu}(p) {V}_{\alpha\mu\nu} (q,r,p) &=& 
P^{\mu'\mu}(r) P^{\nu'\nu}(p) {U}_{\alpha\mu\nu} (q,r,p)  
\nonumber\\
&=& 
P^{\mu'\mu}(r) P^{\nu'\nu}(p) q_{\alpha}[{V}_1 (q,r,p) g_{\mu\nu} + {V}_2 (q,r,p) q_{\mu}q_{\nu}] \,,
\label{Vbarproj}
\eea
so that the only relevant form factors are ${V}_1$ and ${V}_2$. 

At this point we can make the nonperturbative pole  
manifest, and cast ${U}_{\alpha\mu\nu}(q,r,p)$ in the form of Fig.~\ref{Uexpansion}, 
by setting
\be
{U}_{\alpha\mu\nu}(q,r,p) = I_{\alpha}(q)\left(\frac{i}{q^2} \right) \ff_{\mu\nu}(q,r,p) \,, 
\label{VwB}
\ee
where the nonperturbative quantity 
\be
\ff_{\mu\nu}(q,r,p) = B_1g_{\mu\nu}+B_2q_{\mu}q_{\nu}+B_3p_{\mu}p_{\nu}+B_4r_{\mu}q_{\nu}+B_5r_{\mu}p_{\nu} \,,
\ee
is the effective vertex (or ``proper vertex function''~\cite{Jackiw:1973ha})
describing the interaction between 
the massless excitation and two gluons. In the standard language used in bound-state physics, 
$\ff_{\mu\nu}(q,r,p)$ represents the ``bound-state wave function'' (or 
``BS wave function'')  
of the two-gluon bound-state shown in $(b_3)$ of Fig.~\ref{Fullkernel}; as we will see in Section \ref{bse},  
$\ff_{\mu\nu}$ satisfies a (homogeneous) BSE. 
In addition, ${i}/{q^2}$ is the 
propagator of the scalar massless excitation. Finally,   
$I_{\alpha}(q)$ is the (nonperturbative) transition 
amplitude introduced in Fig.~\ref{Uexpansion}, allowing the 
mixing between a gluon and the massless excitation. Note that this latter function is 
universal, in the sense that it enters not only 
in the pole part $V$ associated with the three-gluon vertex, 
but rather 
in all possible such pole parts associated with all other vertices, 
such as the four-gluon vertex, the gluon-ghost-ghost vertex, etc (see panel C in 
Fig.~(\ref{Uexpansion})).

Evidently, by Lorentz invariance,  
\be
I_{\alpha}(q) = q_{\alpha} I(q)\,,
\label{qI}
\ee 
and the scalar cofactor, to be referred to as the ``transition function'', is simply given by 
\be
I(q) = \frac{q^{\alpha}{I}_{\alpha}(q)}{q^2} \,,
\label{trf}
\ee 
so that
\be
{V}_j (q,r,p)  = I(q) \left(\frac{i}{q^2} \right) \ff_j(q,r,p) \, ; \quad j=1,\dots,5 \,. 
\label{pole}
\ee

Note that, due to Bose symmetry (already at the level of $V$) 
with respect to the interchange 
$\mu \leftrightarrow \nu$ and  $p \leftrightarrow r$, we must have
\be
\ff_{1,2}(q,r,p) = - \ff_{1,2}(q,p,r)\,,
\label{anti}
\ee
which implies that 
\be
\ff_{1,2}(0,-p,p) = 0 \,.
\label{ant0}
\ee

Finally, in principle, all other elementary vertices of the theory may also develop pole parts, 
which will play a role completely analogous to that of $V_{\alpha\mu\nu}$ in maintaining the 
corresponding STIs in the presence of a gluon mass. Specifically, in the absence of quarks, 
the remaining vertices are the gluon-ghost-ghost vertex, $\bqq_{\alpha}$,  
the four-gluon vertex $\bqq_{\alpha\mu\nu\rho}$, 
and the  gluon-gluon-ghost-ghost vertex $\bqq_{\alpha\mu}$, which is particular to 
the PT-BFM formulation. 
The parts of their pole vertices containing the $1/q^2$,  denoted by 
$U_{\alpha}$, $U_{\alpha\mu\nu\rho}$, and $U_{\alpha\mu}$, respectively, will all assume the 
common form 
\be 
U_{\alpha\{\dots\}} = I_{\alpha} \left (\frac{i}{q^2}\right) B_{\{\dots\}} \,,
\label{restU}
\ee 
where the various  $B_{\{\dots\}}$ are shown in panel C of Fig.~\ref{Uexpansion}. 

\subsection{An exact relation}

The WI of Eq~(\ref{winp}) furnishes an  
exact relation between the dynamical gluon mass, the transition amplitude at zero momentum transfer, 
and the form factor $B_1$. 
Specifically, contracting both sides of the WI with 
two transverse projectors, one obtains,
\be
P^{\mu'\mu}(r) P^{\nu'\nu}(p) q^\alpha V_{\alpha\mu\nu}(q,r,p) = [m^2(r)-m^2(p)] P^{\mu'}_\sigma(r) P^{\sigma\nu'}(p) \,.
\ee
On the other hand, contracting the full expansion of the vertex (\ref{gendec}) by these transverse projectors
and then contracting the result with the momentum of the background leg, we get 
\be
q^\alpha P^{\mu'\mu}(r) P^{\nu'\nu}(p) V_{\alpha\mu\nu}(q,r,p) = iI(q)[B_1g_{\mu\nu}+B_2q_\mu q_\nu] P^{\mu'\mu}(r) P^{\nu'\nu}(p) \,,
\ee
where the relation of Eq~(\ref{pole}) has been used. Thus, equating both results, one arrives at 
\be
i I(q) B_1(q,r,p) = m^2(r)-m^2(p) \,, \,\,\,\,\,\,\,\,\,\, B_2(q,r,p) = 0 . 
\label{factor1}
\ee
The above relations, together  with those of Eq.~(\ref{pole}),
determine exactly the form factors $V_1$ and $V_2$ of the vertex $V_{\alpha\mu\nu}$, namely
\be
V_1(q,r,p) = \frac{m^2(r)-m^2(p)}{q^2} \,, \,\,\,\,\,\,\,\,\,\,  V_2(q,r,p) = 0 . 
\ee

We will now carry out the Taylor expansion of both sides of Eq~(\ref{factor1}) in the limit \mbox{$q\rightarrow0$}.
To that end, let consider the Taylor expansion of a 
function \mbox{$f(q,r,p)$} around \mbox{$q=0$} (and \mbox{$r=-p$}). In general we have 
\begin{equation}
f(q,-p-q,p)=f(-p,p)+ [2(q\cdot p) + q^2 ]f'(-p,p)+2(q\cdot p)^2 f''(-p,p)+{\cal O}(q^3)\,,
\label{Taylor}
\end{equation}
where the prime denotes differentiation with respect to $(p+q)^2$ and subsequently taking the 
limit $q\to 0$, i.e. 
\begin{equation}
f'(-p,p) \equiv \, \lim_{q\to 0} \left\{ \frac{\partial f(q,-p-q,p)}{\partial\, (p+q)^2} \right\} \,.
\label{Der}
\end{equation}
Now, if the function is antisymmetric under $p \leftrightarrow r$, as happens with 
the form factors $\ff_{1,2}$, then $f(-p,p) = 0$; thus, for the case of the form factors 
in question, the Taylor expansion is ($i=1,2$)
\begin{equation}
\ff_i(q,-p-q,p)= [2(q\cdot p) + q^2] \ff'_i(-p,p) + 2(q\cdot p)^2 \ff''_i(-p,p)  +  {\cal O}(q^3) \,.
\label{TaylorB}
\end{equation}

Using Eq~(\ref{TaylorB}), and the corresponding expansion for the rhs, 
\be
m^2(r)-m^2(p) = m^2(q+p)-m^2(p) = 2(q\cdot p)[m^2(p)]' + {\cal O}(q^2) \quad,
\ee
assuming that the $I(0)$ is finite,
and equating the coefficients in front of $(q\cdot p)$, we arrive at  (Minkowski space)
\be
[m^2(p)]' =  i I(0) B'_1(p) \,. 
\label{massrelation}
\ee
We emphasize that this is an exact relation,
whose derivation relies  only on the 
WI and  Bose-symmetry that $V_{\alpha\mu\nu}(q,r,p )$ 
satisfies, as captured by Eq.~(\ref{winp}) and Eq.~(\ref{ant0}), respectively. 
The Euclidean version of  Eq.~(\ref{massrelation}) is given in  Eq.~(\ref{massrelationeuc}).                      


\subsection{``One-loop dressed'' approximation for the transition function}

We will next approximate the transition amplitude $I_{\alpha}(q)$, connecting the gluon with the 
massless excitation,    
by considering only diagram $(d_1)$ in Fig.~\ref{Uexpansion}, corresponding to the 
gluonic ``one-loop dressed'' approximation;  
we will denote the resulting expression by ${\bar I}_{\alpha}(q)$.
 
In the Landau gauge, ${\bar I}_{\alpha}(q)$ is given by 
\be
{\bar I}_{\alpha}(q) =  \frac{1}{2} C_A 
\int_k \Delta(k) \Delta(k+q) \Gamma_{\alpha\beta\lambda} P^{\lambda\mu}(k) P^{\beta\nu}(k+q) \ff_{\mu\nu}(-q,-k,k+q)\,,
\label{Iappr1}
\ee
where the origin of the factor $1/2$ is combinatoric, 
and 
$\Gamma_{\alpha\beta\lambda}$ is the standard three-gluon vertex at tree-level,  
\be 
\Gamma_{\alpha\mu\nu}(q,r,p) = 
g_{\mu\nu}(r-p)_\alpha  +  g_{\alpha\nu}(p-q)_\mu+g_{\alpha\mu}(q-r)_\nu \,.
\ee
To determine the corresponding transition function from Eq.~(\ref{trf}),   
use that 
\begin{equation}
q^\alpha\Gamma_{\alpha\beta\lambda}(q,-k-q,k)= [k^2 - (k+q)^2]g_{\beta\lambda} + 
[(k+q)_{\beta}(k+q)_{\lambda}  - k_{\beta}k_{\lambda}]\,,
\label{WItree}
\end{equation}
to write  
\be
{\bar I}(q) = -\frac{C_A}{2 q^2}
\int_k  [k^2 - (k+q)^2] \Delta(k)\Delta(q+k) P_{\beta}^{\mu}(k)P^{\beta\nu}(k+q) \,
\ff_{\nu\mu}(-q,k+q,-k) \,.
\label{Iq1}
\ee
In the last step we have used the property of Eq.~(\ref{anti}) in order to 
interchange the arguments of $\ff_{\nu\mu}$, so that the Taylor expansion of Eq.~(\ref{TaylorB}) 
may be applied directly; this accounts for the additional minus sign.  
Then, after the shift $k+q\to k$, and further use of   Eq.~(\ref{anti}),
$\bar I (q)$ becomes   
\be
{\bar I}(q) = -\frac{C_A}{q^2} \int_k  k^2 \Delta(k) \Delta(k+q) P_{\beta}^{\mu}(k) P^{\beta\nu}(k+q) \,
[\ff_1 g_{\mu\nu} + \ff_2 q_{\mu}q_{\nu}]\,.
\label{Iq12}
\ee
To obtain the limit of $\bar I (q)$ as $q^2\rightarrow0$, we will employ 
Eq.~(\ref{TaylorB}) for $\ff_1$ and $\ff_2$, as well as 
\begin{equation}
\Delta(k+q)=\Delta(k)+ [2(q\cdot k) + q^2 ]\Delta'(k)+2(q\cdot k)^2 \Delta''(k)+{\cal O}(q^3) \,.
\label{Taylorgluon}
\end{equation}
Observe that only  the zeroth order term of $P_{\mu\nu}(k+q)$, namely $P_{\mu\nu}(k)$,   
contributes in this expansion.
Then, using spherical coordinates to write $(q\cdot k)^2 = q^2k^2\cos^2\theta$, and the integral 
\begin{equation}
\int_k f(k)\cos^2\theta = \frac{1}{d}\int_k f(k) \,,
\end{equation}
the $\bar I(q)$ in Eq.~(\ref{Iq12}) becomes in the limit \mbox{$q^2\rightarrow0$} (in $d=4$)
\begin{equation}
{\bar I}(0)=-3C_A
\bigg\{\int_k k^2\Delta^2(k)\ff'_1(k)+ \frac{1}{2}\int_k k^4 \frac{\partial}{\partial k^2} [\Delta^2(k)\ff'_1(k)] \bigg\} \,.
\end{equation}
Then, partial integration yields 
\begin{equation}
\int_k k^4\frac{\partial}{\partial k^2}[\Delta^2(k)\ff'_1(k)]=-3\int_k k^2\Delta^2(k)\ff'_1(k)\,,
\end{equation}
and finally one arrives at (Minkowski space) 
\begin{equation}
{\bar I}(0) = \frac{3}{2}C_A\int_k k^2\Delta^2(k)\ff'_1(k)\,.
\label{zero}
\end{equation}
The Euclidean version of this equation, Eq.~(\ref{i0_eu}), will be used in Section \ref{ns}. 

We end this subsection with a comment on the dimensionality  of the various form factors. 
The vertex $V_{\alpha\mu\nu}$ 
has dimension $[m]$, and so 
$V_1$, $V_2$ and $V_3$ are dimensionless, while the remaining form factors have dimension $[m]^{-2}$. 
The integral ${\bar I}(q)$ has the same dimension as $B_1$, and as a result,  
in order to keep $V_1$ dimensionless, $B_1$ must have dimensions of $[m]$.

\subsection{Relating the gluon mass to the transition function}

In this subsection we show how the vertex $V$ gives rise to a gluon mass 
when inserted into the corresponding SDE. We will restrict ourselves to the 
two diagrams shown in Fig.~\ref{gSDE}, and 
will finally express $m^2(0)$ exclusively in terms of $\bar{I}(0)$, which, in turn, depends on the 
existence of $B_{\mu\nu}$ through Eq.~(\ref{zero}). 

In the PT-BFM scheme, the SDE of the gluon propagator 
in the Landau gauge assumes the form 
\be
\Delta^{-1}(q^2){P}_{\mu\nu}(q) = 
\frac{q^2 {P}_{\mu\nu}(q) + i\,\Pi_{\mu\nu}(q) }{[1+G(q^2)]^2} \,.
\label{sde}
\ee
The most straightforward way to relate the gluon mass to the transition function $\bar{I}$
is to identify, on both sides of (\ref{sde}),  
the co-factors of the tensorial structure 
$q_{\mu}q_{\nu}/q^2$ which survive the limit $q^2\to 0$, and then set them equal to each other.
Making the usual identification (in Minkowski space)  $\Delta^{-1}(0) = - m^2(0)$, 
it is clear that lhs of  (\ref{sde}) furnishes simply 
\be
{\rm lhs}|_{\frac{q_{\mu}q_{\nu}}{q^2}} = m^2(0) \,.
\label{lhs}
\ee

\begin{figure}[t]
\center{\includegraphics[scale=0.6]{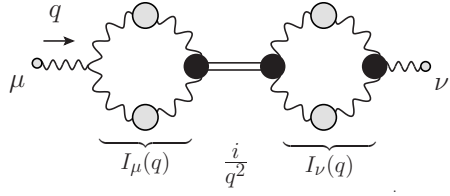}}
\caption{The ``squared'' diagram.}
\label{Square}
\end{figure}

It is relatively straightforward to recognize that 
the analogous contribution from the rhs comes from the standard ``squared'' diagram,  
shown in Fig.~\ref{Square}. Specifically, the starting expression is 
\be
\Pi^{\mu\nu}(q) = \frac{1}{2} g^2 C_A \int_k
\Gamma^{\mu}_{\alpha\beta}P^{\alpha\rho}(k)P^{\beta\sigma}(k+q)
[\bqq+V]^\nu_{\rho\sigma} \Delta(k) \Delta(k+q) + \cdots \,,
\ee
where, as explained earlier, the (all order) vertex $\bqq$ has been replaced by $\bqq+V$, 
and the ellipses denote terms that, in the kinematic limit considered, do not contribute to the 
specific structure of interest.

The relevant contribution originates from the part containing the vertex $V$, 
to be denoted by $\Pi^{\mu\nu}(q)|_{\chic V}$; it is 
represented by the diagram in Fig.~\ref{Square}. 
In particular, by virtue of Eq.~(\ref{Vbarproj}), we have   
\bea
\Pi^{\mu\nu}(q)|_{\chic V}  &=& 
\frac{1}{2} g^2 C_A \int_k \Delta(q+k)\Delta(k) 
\Gamma^{\mu}_{\alpha\beta}P^{\alpha\rho}(k)P^{\beta\sigma}(k+q) U^\nu_{\rho\sigma}
\nonumber\\
&=& 
g^2 \left\{\frac{1}{2} C_A \int_k \Delta(q+k)\Delta(k) 
\Gamma^{\mu}_{\alpha\beta}P^{\alpha\rho}(k)P^{\beta\sigma}(k+q)
\ff_{\rho\sigma} \right\} \left(\frac{i}{q^2} \right){\bar I}_{\nu}(q) 
\nonumber\\
&=& i \frac{q^{\mu}q^{\nu}}{q^2} \,g^2  {\bar I}^2 (q)\,,
\label{A1v}
\eea
where in the second line we have used Eq.~(\ref{VwB}) [ with $I_{\nu}(q) \to {\bar I}_{\nu}(q)$], 
and Eq.~(\ref{Iappr1}) in the third line. 

Thus, using the fact that,  since $L(0)=0$~\cite{Aguilar:2009nf}, from the identity 
of Eq.~(\ref{funrel}) we have that  $1+G(0) = F^{-1}(0)$, then  
the rhs of (\ref{sde}) becomes 
\be
{\rm rhs}|_{\frac{q_{\mu}q_{\nu}}{q^2}} = - g^2 F^2(0) {\bar I}^2 (0) \,.
\label{rhs}
\ee
We next go to Euclidean space, following the usual rules, and noticing that, due to the 
change $\int_k\!= i \!\int_{k_{\chic E}}$ we have  
${\bar I}^2 (0) \to -{\bar I}^2_{\chic E} (0)$; so, equating (\ref{lhs}) and (\ref{rhs}) 
we obtain  (suppressing the index ``E'') 
\be
m^2(0)= g^2 F^2(0) {\bar I}^2 (0) \,.
\label{me-geneuc}
\ee
Note that the $m^2(0)$ so obtained is positive-definite. 
We emphasize that the relation of Eq.~(\ref{me-geneuc}) constitutes the (gluonic) ``one-loop dressed''  
approximation of the complete relation; indeed, both the SDE 
used as starting point
as well as the expression for ${\bar I}$ are precisely the corresponding ``one-loop dressed''
contributions, containing gluons (but not ghosts).

Finally, let us consider the exact relation~\cite{Aguilar:2011ux} 
\be
\widehat{m}^2(q^2)=[1+G(q^2)]^2 m^2(q^2),
\label{mhat-m}
\ee
expressing the dynamical mass $m(q^2)$ of the standard gluon propagator $\Delta(q^2)$
in terms of the corresponding mass, $\widehat{m}(q^2)$, of the PT-BFM gluon propagator
[usually denoted by $\widehat\Delta(q^2)$] in the same gauge 
[in this case, in the Landau gauge].  
At $q^2=0$ this relation reduces to $m^2(0) = \widehat{m}^2(0) F^2(0)$, so that Eq.~(\ref{me-geneuc}) 
may be alternatively written as 
\be
{\widehat m}^2(0) =  g^2 {\bar I}^2 (0) \,.
\label{PTresfin}
\ee 
Interestingly enough, when written in this form, the mass formula derived from our SDE analysis 
coincides with the one obtained for the photon mass
in the Abelian model of Jackiw and Johnson (Eq.~(2.12) in~\cite{Jackiw:1973tr}).   
In addition,  this last form facilitates the demonstration of the decoupling of 
the massless excitation from the on-shell four-gluon amplitude (see Section \ref{dec}).   

In principle, the analysis presented above may be extended  
to include the rest of the graphs contributing to the gluon SDE, 
invoking the corresponding pole parts of the remaining vertices; however,  
this lies beyond the scope of the  present work.

\section{\label{bse}BS equation for the bound-state wave function}

As has become clear in the previous section, the gauge boson (gluon) mass 
is inextricably connected to the existence of the quantity $\ff'_1$.
Indeed, if $\ff'_1$ were to vanish, then, by virtue of  (\ref{zero})
so would ${\bar I}(0)$, and therefore, through (\ref{me-geneuc}) we would 
obtain a vanishing $m^2(0)$. Thus, the existence of $\ff'_1$ is of paramount 
importance for the mass generation mechanism envisaged here; essentially, the question 
boils down to whether or not   
the dynamical formation of a massless 
bound-state excitation of the type postulated above is possible.   
As is well-known, in order to establish the 
existence of such a bound state one must 
{\bf (i)} derive the appropriate BSE for the 
corresponding bound-state wave function, $B_{\mu\nu}$,  
(or, in this case, its derivative),   
and {\bf (ii)} find non-trivial solutions for this integral equation.

To be sure, this dynamical equation will be derived under certain simplifying assumptions, 
which will be further refined in order to obtain numerical solutions. We emphasize, therefore,  
that the analysis presented here is meant to provide preliminary quantitative evidence    
for the realization of the dynamical scenario considered, 
but cannot be considered as a conclusive demonstration.

\begin{figure}[!t]
\center{\includegraphics[scale=0.6]{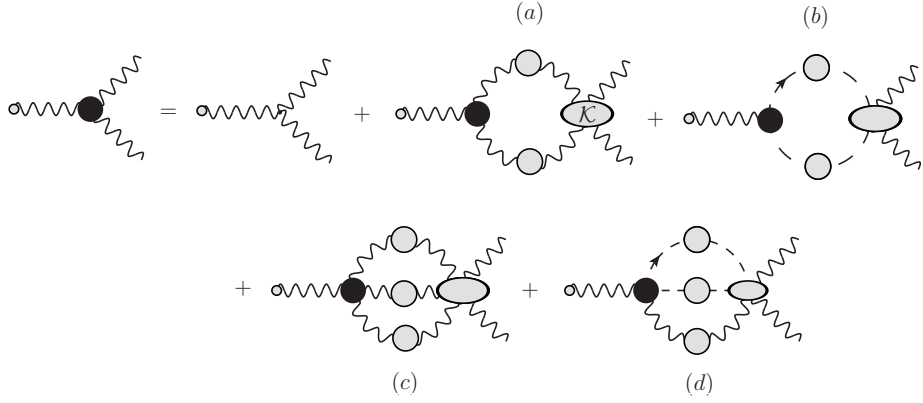}}
\caption{The complete BSE for the full three gluon vertex $\NV_{\alpha\mu\nu}(q,r,p)$.}
\label{Fig6}
\end{figure}

\begin{figure}[!b]
\center{\includegraphics[scale=0.55]{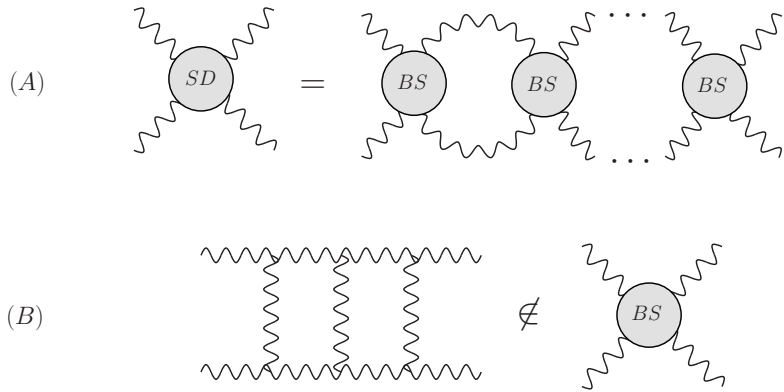}}
\caption{{\bf(A)} Schematic relation between the SDE and BSE kernels.
{\bf(B)} Example of a diagram not contained in the corresponding  
BSE kernel, in order to avoid over counting.}
\label{DSBS}
\end{figure}

The starting point is the BSE for the vertex $\NV_{\alpha\mu\nu}(q,r,p)$, shown in Fig.~\ref{Fig6}.
Note that, unlike the corresponding SDE of  Fig.~\ref{FullvertexSDE}, the vertices where the 
background gluon is entering (carrying momentum $q$) are now fully dressed.
 As a consequence, the corresponding multiparticle kernels appearing in   Fig.~\ref{Fig6} 
are different from those 
of the SDE, as shown in Fig.~\ref{DSBS}.

\begin{figure}[!t]
\center{\includegraphics[scale=0.5]{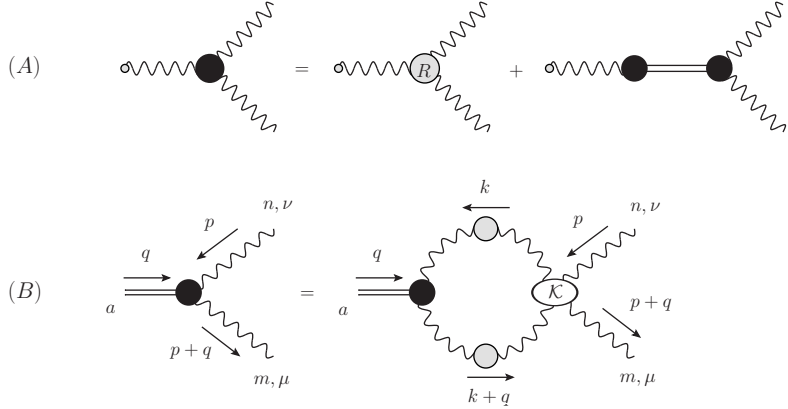}}
\caption{{\bf(A)} The separation of the vertex in regular and pole parts.
{\bf(B)} The BSE for the bound-state wave function $B_{\mu\nu}$.}
\label{RegPole}
\end{figure}

The general methodology of how to isolate from the BSE shown in Fig.~\ref{Fig6} 
the corresponding dynamical equation for 
the quantity $B_{\mu\nu}$ has been explained in~\cite{Jackiw:1973ha,Poggio:1974qs}. 
Specifically, one separates on both sides of the BSE equation 
each vertex (black circle) into two parts,  
a ``regular'' part and another containing a pole $1/q^2$; this separation 
is shown schematically in Fig.~\ref{RegPole}. 
Then, the BSE for $B_{\mu\nu}(q,r,p)$
is obtained simply by equating the pole parts on both sides. Of course, for the case we consider 
the full implementation of this general procedure would lead to 
a very complicated structure, because, in principle, 
all fully dressed vertices appearing on the rhs of Fig.~\ref{Fig6} may contain pole parts 
[i.e., not just the three-gluon vertex of (a) but also those  in (b), (c), and (d)].  
Thus, one would be led to an equation, 
whose lhs would consist of $B_{\mu\nu}$, but whose rhs would contain the $B_{\mu\nu}$
together with all other similar vertices, denoted by $B_{\{\dots\}}$ in Eq.~(\ref{restU}). 
Therefore, this equation must be supplemented by a set of analogous equations, obtained from 
the BSEs of all other vertices appearing on the rhs of Fig.~\ref{Fig6} [i.e., those in (b), (c), (d) ]. 
So, if all vertices involved contain a pole part, one would arrive 
at a system of several coupled integral equations, containing complicated combinations of the 
numerous form factors composing these vertices (see, for example, Fig.~11 in~\cite{Poggio:1974qs}).

It is clear that for practical purposes 
the above procedure must be simplified to something more manageable. 
To that end, we will only consider graph (a) on the rhs of Fig.~\ref{Fig6}, thus reducing the 
problem to the treatment of a single integral equation.

Specifically, 
the BSE  
for $B_{\mu\nu}$ is given by [see Fig.~\ref{Fig6}]
\begin{equation}
B_{\mu\nu}^{amn} = \int_k B_{\alpha\beta}^{abc}\Delta^{\alpha\rho}_{br}(k+q)\Delta^{\beta\sigma}_{cs}(k){\cal K}_{\sigma\nu\mu\rho}^{snmr} \,.
\end{equation}

\begin{figure}[!t]
\center{\includegraphics[scale=0.5]{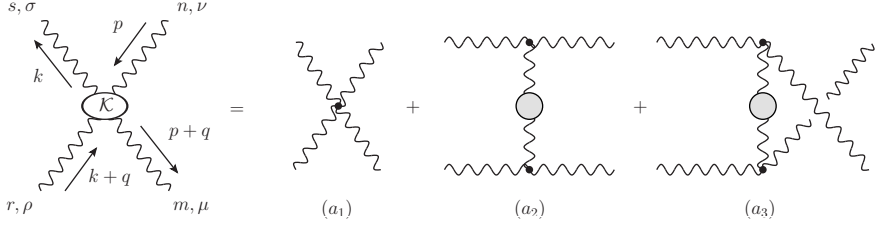}}
\caption{The Feynman diagrams considered for the BS kernel. 
The interaction vertices are approximated by their tree level values, 
while the internal gluon propagators are fully dressed.}
\label{Fig9}
\end{figure}

In addition, we will approximate the  four-gluon BS kernel ${\cal K}$ by the 
lowest-order set of diagrams shown in Fig.~\ref{Fig9}, where the vertices are bare, while 
the internal gluon propagators are fully dressed.

To proceed further, 
observe that the diagram $(a_1)$ does not 
contribute to the BSE, 
because the color structure of the tree-level four-gluon vertex
vanishes when contracted with the color factor $f^{abc}$ of the $B_{\alpha\beta}$.
Diagrams $(a_2)$ and $(a_3)$ are equal, and are multiplied by a  Bose symmetry factor of $1/2$.  
So, in this approximation, the BS kernel is given by
\begin{equation}
{\cal K}_{\sigma\nu\mu\rho}^{snmr}(-k,p,-p-q,k+q)=-ig^2f^{sne}f^{emr}
\Gamma^{(0)}_{\sigma\gamma\nu}\Delta^{\gamma\lambda}(k-p)\Gamma^{(0)}_{\mu\lambda\rho} \,,
\end{equation}
where $\Gamma^{(0)}$ is the tree-level value of the three gluon vertex.
So, using this kernel 
and setting the gluon propagators in the Landau gauge, 
the BSE becomes 
\begin{equation}
B_{\mu\nu} = -2\pi i\alpha_sC_A\int_k B_{\alpha\beta}\Delta(k+q)\Delta(k)\Delta(k-p)P^{\alpha\rho}(k+q)P^{\beta\sigma}(k)P^{\gamma\lambda}(k-p)\Gamma^{(0)}_{\sigma\gamma\nu}\Gamma^{(0)}_{\mu\lambda\rho} \,,
\label{generalBS}
\end{equation}
where we have cancelled out a color factor $f^{abc}$ from both sides.

Let us focus on the lhs of Eq.~(\ref{generalBS}). 
Using the Taylor expansion in Eq.~(\ref{TaylorB}), the fact that $B_2=0$ [see Eq.~(\ref{factor1})],  
and multiplying by a transverse projector we obtain,
\begin{equation}
P^{\mu\nu}(p) B_{\mu\nu} = 6 (q\cdot p) B_1'(p) + {\cal O}(q^2) \quad.
\end{equation}
Next, let us denote by $[{\rm rhs}]_{\mu\nu}$ the rhs of Eq.~(\ref{generalBS}).  
Inserting the bare value for the three gluon vertices, multiplying by the transverse projector,   
and  using the Taylor expansions in Eq.~(\ref{TaylorB}) and (\ref{Taylorgluon}), after standard  
manipulations one obtains the result
\begin{equation}
P^{\mu\nu}(p) \, [{\rm rhs}]_{\mu\nu} = 
-4\pi i\alpha_sC_A (q\cdot p)\int_k B_1'(k)\Delta^2(k)\Delta(k-p){\cal N}(p,k) + 
{\cal O}(q^2) \,,
\end{equation}
where we have defined the kernel 
\begin{equation}
{\cal N}(p,k)=\frac{4(p\cdot k)[p^2k^2-(p\cdot k)^2]}{p^4k^2(k-p)^2}[8p^2k^2-6(pk)(p^2+k^2)+3(p^4+k^4)+(pk)^2]\,.
\end{equation}
Thus, equating the lhs with the rhs, we derive the following BSE for the derivative of the form factor 
that appears in the mass relation Eq.~(\ref{massrelation}),
\begin{eqnarray}
B'_1(p)=-\frac{2\pi i}{3}\alpha_s C_A \int_k B'_1(k)\Delta^2(k)\Delta(k-p){\cal N}(p,k) \,.
\end{eqnarray}

Going to Euclidean space, 
we define
\begin{equation}
x\equiv p^2 \,; \quad y\equiv k^2 \, ; \quad z\equiv (p+k)^2 \,,
\end{equation}
and write the Euclidean integration measure in spherical coordinates,
\begin{equation}
\int\frac{d^4k_E}{(2\pi)^4} = \frac{1}{(2\pi)^3}\int_0^\infty dyy\int_0^\pi d\theta \sin^2\theta \,,
\end{equation}
so that the BSE becomes
\begin{eqnarray}\label{34}
B'_1(x) &=& -\frac{\alpha_s C_A}{12\pi^2}\int_0^\infty dy yB'_1(y)\Delta^2(y)\nonumber \\
&& \sqrt{\frac{y}{x}}\int_0^\pi d\theta\sin^4\theta\cos\theta\bigg[z+10(x+y)+\frac{1}{z}(x^2+y^2+10xy)\bigg]\Delta(z)\,.
\label{euclideanBS}
\end{eqnarray}

In spherical coordinates we have that $z=x+y+2\sqrt{xy}\cos\theta$. So, around $x=0$,
\begin{equation}
\frac{1}{z}=\frac{1}{x+y}\bigg[1-2\frac{\sqrt{xy}}{x+y}\cos\theta\bigg]\,,
\end{equation}
and using the Taylor expansion for the gluon propagator $\Delta(z)$, 
the limit $x\rightarrow0$ can be taken in the BSE,  giving the value
\begin{equation}
B'_1(0)=\lim_{x\rightarrow0}B'_1(x)=-\frac{\alpha_s C_A}{8\pi}\int_0^\infty dyy^3
B'_1(y)\Delta^2(y)\Delta'(y)\,.
\label{zerovalue}
\end{equation}

Let us finally 
implement an additional simplification to Eq.~(\ref{euclideanBS}), 
which will allow us to carry out the angular integration 
exactly, thus reducing the problem to the solution of 
a one-dimensional  integral equation.  
Specifically, the simplification consists in approximating the gluon propagator $\Delta(z)$ 
appearing in the BSE of (\ref{euclideanBS}) [but not the $\Delta^2(y)$]
by its tree level value, that is, $\Delta(z)=1/z$. Then, 
with the aid of the angular integrals,
\bea
\sqrt{\frac{y}{x}}\int_0^\pi d\theta\frac{\sin^4\theta\cos\theta}{z} 
&=& \frac{\pi}{16x}\left[\frac{y}{x^2}(y-2x)\Theta(x-y)+\frac{x}{y^2}(x-2y)\Theta(y-x)\right]\,,
\nonumber \\
\sqrt{\frac{y}{x}}\int_0^\pi d\theta\frac{\sin^4\theta\cos\theta}{z^2} &=& 
-\frac{\pi}{4x}\left[\frac{y}{x^2}\Theta(x-y)+\frac{x}{y^2}\Theta(y-x)\right]\,,
\eea
one brings Eq.~(\ref{euclideanBS}) into the form 
\begin{eqnarray}\label{39}
B'_1(x) &=& \frac{\alpha_s C_A}{24\pi}\bigg\lbrace\int_0^x dyB'_1(y)\Delta^2(y)\frac{y^2}{x}
\bigg(3+\frac{25}{4}\frac{y}{x}-\frac{3}{4}\frac{y^2}{x^2}\bigg)+ \nonumber \\
&& +\int_x^\infty dyB'_1(y)\Delta^2(y)y\bigg(3+\frac{25}{4}\frac{x}{y}-\frac{3}{4}\frac{x^2}{y^2}\bigg)\bigg\rbrace \,.
\label{weakBS}
\end{eqnarray}

The limit $x\rightarrow0$ of this equation 
is given by (the change of variable $y=tx$ may be found useful),
\begin{equation}
B'_1(0)=\frac{\alpha_s C_A}{8\pi}\int_0^\infty dy yB'_1(y)\Delta^2(y)\,. 
\label{zerolimit}
\end{equation}
Note that this result coincides, as it should, with that obtained from Eq.~(\ref{zerovalue}) 
after setting $\Delta'(y) = -1/y^2$, 
namely the derivative of the tree-level propagator.

\section{\label{ns}Numerical solutions and existence of a bound-state}

In this section we will carry out a detailed numerical analysis 
of the integral equation obtained in the previous section, namely Eq.~(\ref{weakBS}). 

First of all, let us point out that, despite appearances, 
the integral equation (\ref{weakBS}) is not linear in the unknown 
function $B'_1(x)$. The non-linearity enters through the 
propagator $\Delta(y)$, which depends on the dynamical mass $m^2(y)$ 
through Eq.~(\ref{defm}); as a result, and by virtue of Eq.~(\ref{massrelation}),
which, in Euclidean space reads 
\be
[m^2(y)]' = -I(0) B'_1(y)\,,
\label{massrelationeuc}
\ee
it is clear that $\Delta(y)$ depends on $B'_1(x)$ in a complicated way. 
Specifically, from the two aforementioned equations we have  
\bea 
\Delta^{-1}(y) &=& yJ(y) + m^2(y)\,,
\nonumber \\
m^2(y) &=& m^2(0) -I(0) \int_{0}^{y} dz  B'_1(z) \,.
\label{exeq}
\eea
where $I(0)$ may be approximated by its ``one-loop dressed'' version ${\bar I}(0)$ given in (\ref{zero}), which 
in Euclidean space becomes
\be
{\bar I}(0) = \frac{3C_A}{32\pi^2} \int_0^{\infty} \!\!\!\! dy\, y^2\Delta^2(y)B'_1(y) \,.
\label{i0_eu}
\ee

Then, Eq.~(\ref{weakBS}) must be solved together with the two additional 
relations given in Eq.~(\ref{exeq}), as a non-linear system. 
In addition, one may use Eq.~(\ref{me-geneuc}), 
in order to obtain an (approximate) constraint for $I(0)$.   
Note also that Eq.~(\ref{weakBS}), again due to  Eq.~(\ref{massrelationeuc}), 
may be recast entirely in terms of $m^2(y)$ and its derivative.

For the purposes of the present work we will simplify somewhat the 
procedure described above. Specifically, 
we will present two different approaches, each one particularly suited 
for probing distinct features of Eq.~(\ref{weakBS}) 
and the accompanying Eqs.~(\ref{exeq}).
In particular, we will first study  Eq.~(\ref{weakBS}) 
in isolation, using simple Ans\"atze for $\Delta(y)$. The purpose of this
study is to establish the existence of non-trivial solutions for 
$B'_1$, study their dependence on the value of the strong coupling $\alpha_s$,
and verify the asymptotic behavior predicted by Eq.~(\ref{algeq2}).
Of course, since  $\Delta(y)$ 
at this level is treated as an ``external'' object, the 
homogeneous Eq.~(\ref{weakBS}) 
becomes linear in $B'_1$; as a result, given one solution we obtain a family of such solutions,
through multiplication by any real constant. 
Then, as a second step, we will use the available lattice data 
for the gluon propagator $\Delta(y)$, in order to obtain the corresponding solution 
for $B'_1$. Now, the linearity induced by treating $\Delta(y)$ as an external 
input will be resolved by resorting to  Eq.~(\ref{exeq}) and Eq.~(\ref{me-geneuc}); 
thus, out of the infinite family of solutions only one will be dynamically selected.
These two approaches will be presented in subsections~\ref{bb3} and~\ref{cc3}, while subsection~\ref{asy3} 
deals with the asymptotic behavior of the solutions. 

\subsection{Asymptotic behavior}
\label{asy3}

\begin{figure}[ht]
\center{\includegraphics[scale=0.55]{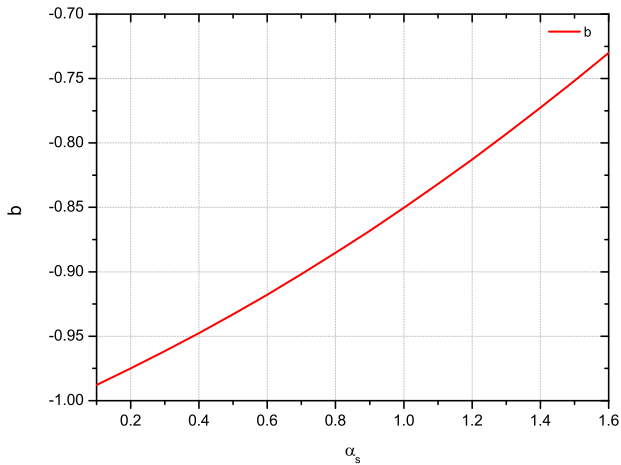}}
\caption{The  physically relevant solution of Eq.~(\ref{poly}).}
\label{exp}
\end{figure}

Before turning to the numerical treatment of Eq.~(\ref{weakBS}), 
it is useful to study the behavior of the solutions  
for asymptotically large values of $x$. Setting $\Delta(y) = 1/y$, it is 
relatively straightforward to establish that 
the equation admits a power-law solution of the form 
$B'_1(x) = A x^{b}$. Specifically, substituting this Ansatz into
the first integral of  Eq.~(\ref{weakBS}), which is the dominant part for large $x$,  
and carrying out the integrations, one arrives at  
the following algebraic equation for $b$, 
\be
\frac{24 \pi}{\alpha_s C_A} = \frac{3}{b+1}
+\frac{25}{4(b+2)}- \frac{3}{4(b+3)}\,,
\label{algeq2}
\ee
together with the restriction $b>-1$, imposed in order to assure convergence 
in the lower ($y=0$) limit of integration. 
Setting $\lambda \equiv 24 \pi/\alpha_s C_A$, one arrives 
at the third-order equation
\be
4\lambda b^3 - (34 -24\lambda)b^2 - (151 -44\lambda)b -141 +24\lambda =0 \,,
\label{poly}
\ee
that may be easily solved;  the solution that satisfies $b>-1$ 
is shown in Fig.~\ref{exp} as a function of $\alpha_s$. 

\subsection{The linearized case:~solutions for various gluon propagators}
\label{bb3}

Next we discuss the numerical solutions for 
Eq.~(\ref{weakBS}) for arbitrary values of $x$. 
Evidently, the main ingredient entering into its kernel is
the nonperturbative gluon propagator, $\Delta(q)$. In order to explore the sensitivity of 
the solutions on the details of $\Delta(q)$, 
we  will employ three infrared-finite forms, to be denoted by $\Delta_1(q)$, $\Delta_2(q)$, 
and $\Delta_3(q)$, focusing on their differences  in the intermediate and asymptotic 
regions of momenta.

\begin{figure}[!t]
\begin{minipage}[b]{0.45\linewidth}
\noindent
\centering
\hspace{-1cm}
\includegraphics[scale=0.55]{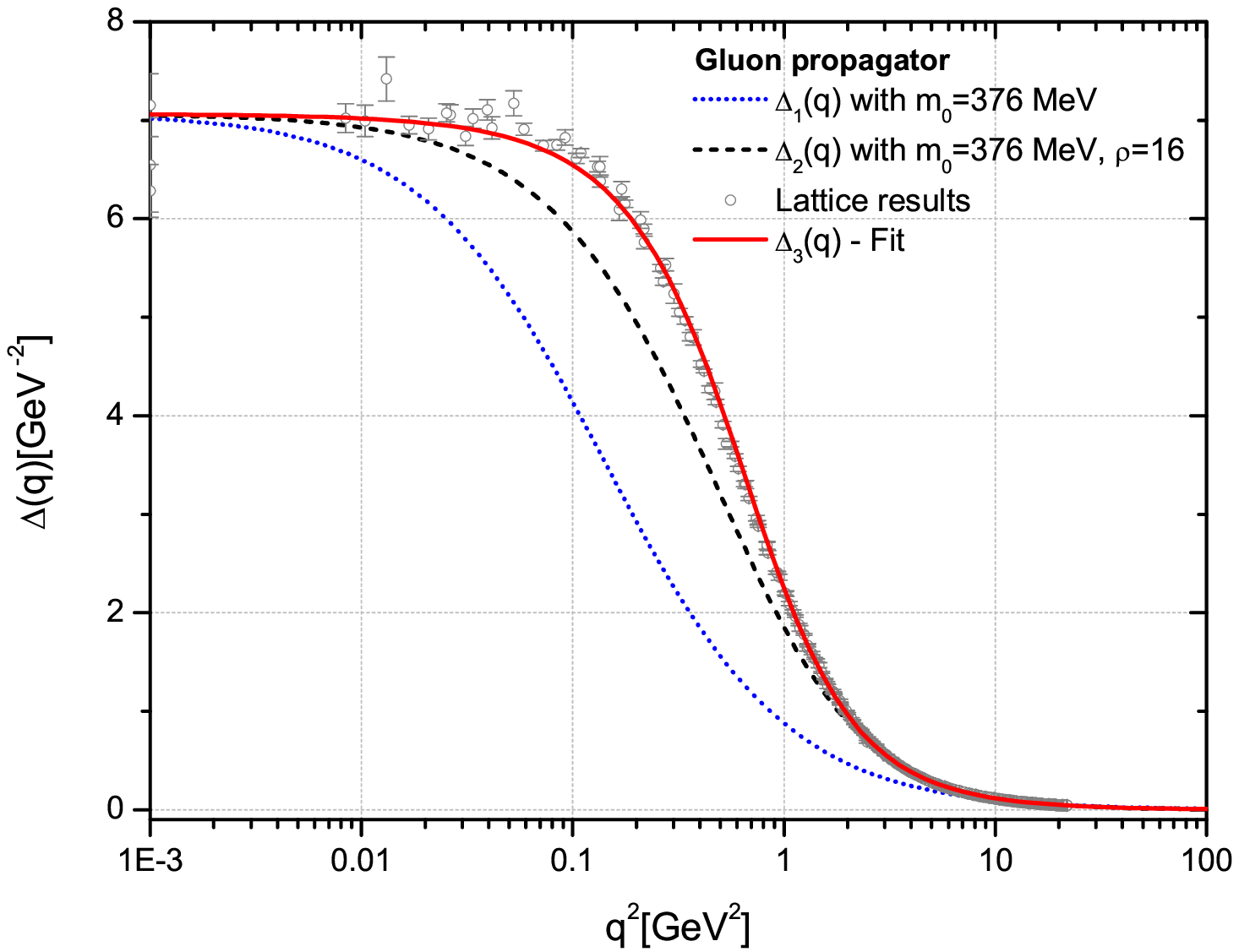}
\end{minipage}
\begin{minipage}[b]{0.50\linewidth}
\hspace{-1.5cm}
\noindent
\includegraphics[scale=0.55]{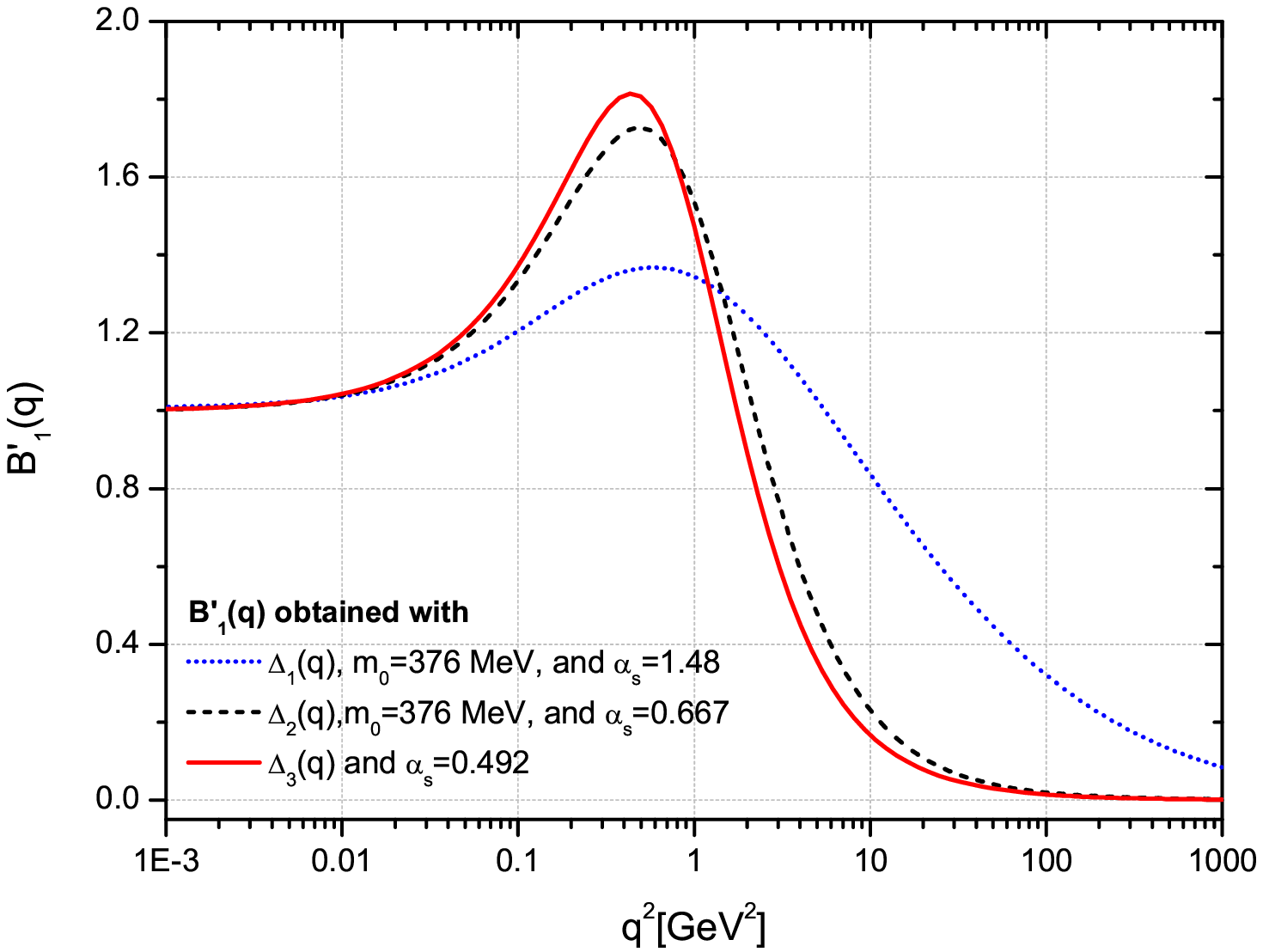}
\end{minipage}
\caption{{\it Left panel}: The three models for the gluon propagator
as function of the momentum $q^2$. The (red) continuous line
is the fit for the lattice gluon propagator given by Eq.~(\ref{gluon3}) when \mbox{$m= 520\,$ MeV},
\mbox{$g_1^2=5.68$}, \mbox{$\rho_1=8.55$}, \mbox{$\rho_2=1.91$}, 
and $\mu=4.3$ GeV; the (black) dashed line
is the model of Eq.~(\ref{gluon2}) with $\rho=16$, $\alpha_s=0.667$ and $m_0=376 \,\mbox{MeV}$, while the (blue) dotted line
represents the massive propagator of Eq.~(\ref{smassive}) when $m_0=376 \,\mbox{MeV}$.  
{\it Right panel}: The corresponding solutions of Eq.~(\ref{weakBS}) obtained with the gluon 
propagators shown on the left panel. The solutions for $B'_1(q)$ are obtained
when we fix the  value of \mbox{$\alpha_s=1.48$}, \mbox{$\alpha_s=0.667$}, and \mbox{$\alpha_s=0.492$} for 
$\Delta_1(q)$, $\Delta_2(q)$, and $\Delta_3(q)$, respectively.}
\label{props}
\end{figure}

Let us start with the simplest such propagator, namely a tree-level massive propagator of the form 
\be
\Delta_1^{-1}(q^2) = q^2 + m^2_0 \,,
\label{smassive}
\ee 
where $m^2_0$ is a hard mass, that will be treated as a free parameter. 
On the left panel of Fig.~\ref{props}, the (blue) dotted curve represents 
$\Delta_1(q^2)$ for $m_0=376 \,\mbox{MeV}$.

The second model is an improved version of the first, where 
we introduce the renormalization-group logarithm next to the momentum $q^2$, more specifically
\be
\Delta_2^{-1}(q^2)=  q^2\left[1+ \frac{13C_{\rm A}g^2}{96\pi^2} 
\ln\left(\frac{q^2 +\rho\,m_0^2}{\mu^2}\right)\right] +  m^2_0 \,.
\label{gluon2}
\ee  
where $\rho$ is an adjustable parameter varying in the range of $\rho \in [2,10]$. Notice that the hard mass $m_0^2$
appearing in the argument of the perturbative logarithm  
acts as an infrared cutoff; so, instead of the logarithm diverging at the Landau pole, it
saturates at a finite value. The (black) dashed line
represents the Eq.~(\ref{gluon2}) when $\rho=16$, $m_0=376 \,\mbox{MeV}$, and  $\mu=4.3$ GeV.

The third model is simply a 
physically motivated fit for the gluon propagator determined
by the large-volume lattice simulations of Ref.~\cite{Bogolubsky:2007ud}, and 
shown on the left panel of Fig.~\ref{props}. 
The lattice data presented there correspond to a $SU(3)$  quenched 
lattice simulation, where $\Delta(q)$ is renormalized at $\mu=4.3$ GeV.  
This gluon propagator 
can be  accurately fitted by the expression (e.g.,\cite{Aguilar:2011ux})
\be
\Delta_3^{-1}(q^2)= m_g^2(q^2) + q^2\left[1+ \frac{13C_{\rm A}g_1^2}{96\pi^2} 
\ln\left(\frac{q^2 +\rho_1\,m_g^2(q^2)}{\mu^2}\right)\right]\,,
\label{gluon3}
\ee  
where  $m_g^2(q^2)$ is a running mass given by
\be
m^2_g(q^2) = \frac{m^4}{q^2 + \rho_2 m^2} \,,
\label{dmass}
\ee
and the values of the fitting parameters are  
\mbox{$m= 520$\,\mbox{MeV}}, \mbox{$g_1^2=5.68$}, \mbox{$\rho_1=8.55$} and, \mbox{$\rho_2=1.91$}.
On the left panel of Fig.~\ref{props}, the (red) continuous line represents
the fit for the lattice gluon propagator given by Eq.~(\ref{gluon3}). Notice that, in all three cases, we 
have fixed the value of $\Delta^{-1}(0)=m_0^2\approx 0.14$.

\begin{figure}[!t]
\begin{center}
\includegraphics[scale=0.55]{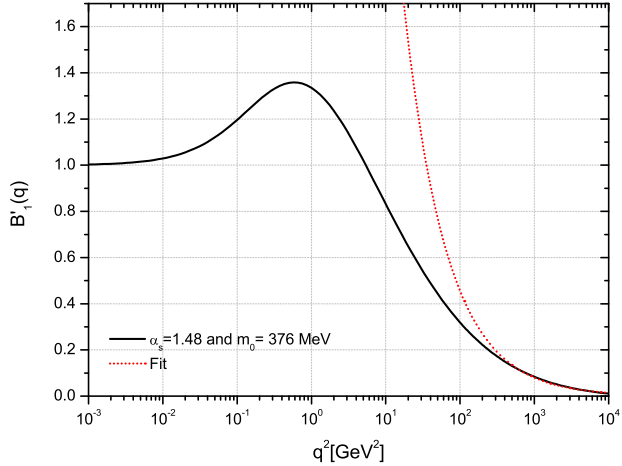}
\end{center}
\vspace{-1.0cm}
\caption{The (black) continuous curve  represents the solution obtained 
from  Eq.~(\ref{weakBS}) using the propagator 
$\Delta_1(q)$ of Eq.~(\ref{smassive}), with $\alpha_s=1.48$ and $m_0=376 \,\mbox{MeV}$. 
The (red) dotted line is the best fit obtained
for the asymptotic behavior of $B'_1(x)$ given by $B'_1(q) = Aq^{2b}$ with $A_1=14.80$ and $b=-0.756$. Notice
that this value is in excellent agreement with 
the power found by the  analytical determination, shown in Fig.~\ref{exp}.}   
\label{asymp}
\end{figure}

Our main findings may be summarized as follows.

{\it (i)} In Fig.~\ref{props}, right panel, we show the solutions 
of Eq.~(\ref{weakBS}) obtained using as input 
the three propagators shown on the left panel.
For the simple massive propagator
of Eq.~(\ref{smassive}), a solution for $B'_1(q)$ is found for \mbox{$\alpha_s=1.48$};
in the case of $\Delta_2(q)$ given by  Eq.~(\ref{gluon2}), a solution 
is obtained when \mbox{$\alpha_s=0.667$}, while for the lattice
propagator $\Delta_3(q)$ of  Eq.~(\ref{gluon3}) a non-trivial 
solution is found when \mbox{$\alpha_s=0.492$}.

{\it (ii)} Note that, due to the fact that Eq.~(\ref{weakBS}) is homogeneous and (effectively) linear,
if $B'_1(q)$ is a solution then the function $cB'_1(q)$ is also a solution, 
for any real constant $c$. Therefore,
the solutions shown on the right panel of Fig.~\ref{props} corresponds to  
a representative case of a family of  possible solutions, where the constant $c$ was chosen 
such that $B'_1(0)=1$.

{\it (iii)} Another interesting feature of the solutions of Eq.~(\ref{weakBS}) is  
the dependence of the observed peak on the support of the gluon propagator 
in the intermediate region of momenta. Specifically, 
an increase of the support of the gluon propagator in the approximate range (0.3-1) GeV 
results in a more pronounced peak in  $B'_1(q)$.

{\it (iv)} In addition, observe that due to the presence of the 
perturbative logarithm 
in the expression for $\Delta_2(q)$ and $\Delta_3(q)$, the corresponding solutions $B'_1(q)$  
fall off in the ultraviolet region much faster than those obtained using the 
simple $\Delta_1(q)$ of Eq.~(\ref{smassive}). 
In order to check whether the power-law  asymptotic behavior, $B'_1(q) = Aq^{2b}$, determined in our 
previous analysis, is in agreement with our numerical solution, we isolate
in Fig.~\ref{asymp} the solution of $B'_1(q)$ obtained with $\Delta_1(q)$ and
\mbox{$\alpha_s=1.48$} (black  continuous curve) and compare it with  the best fit
obtained for large values of $q^2$ (red dotted curve). 
Indeed,  the asymptotic tail of $B'_1(q)$ falls off as power law
of the type $B'_1(q) = Aq^{2b}$  with $A=14.80$ and $b=-0.756$. Notice that   
the value of  $b$ obtained from the fit is in perfect
agreement with values obtained from  Eq.~(\ref{poly}), shown in Fig.~\ref{exp}.

\begin{figure}[!t]
\begin{minipage}[b]{0.45\linewidth}
\noindent
\centering
\hspace{-1cm}
\includegraphics[scale=0.55]{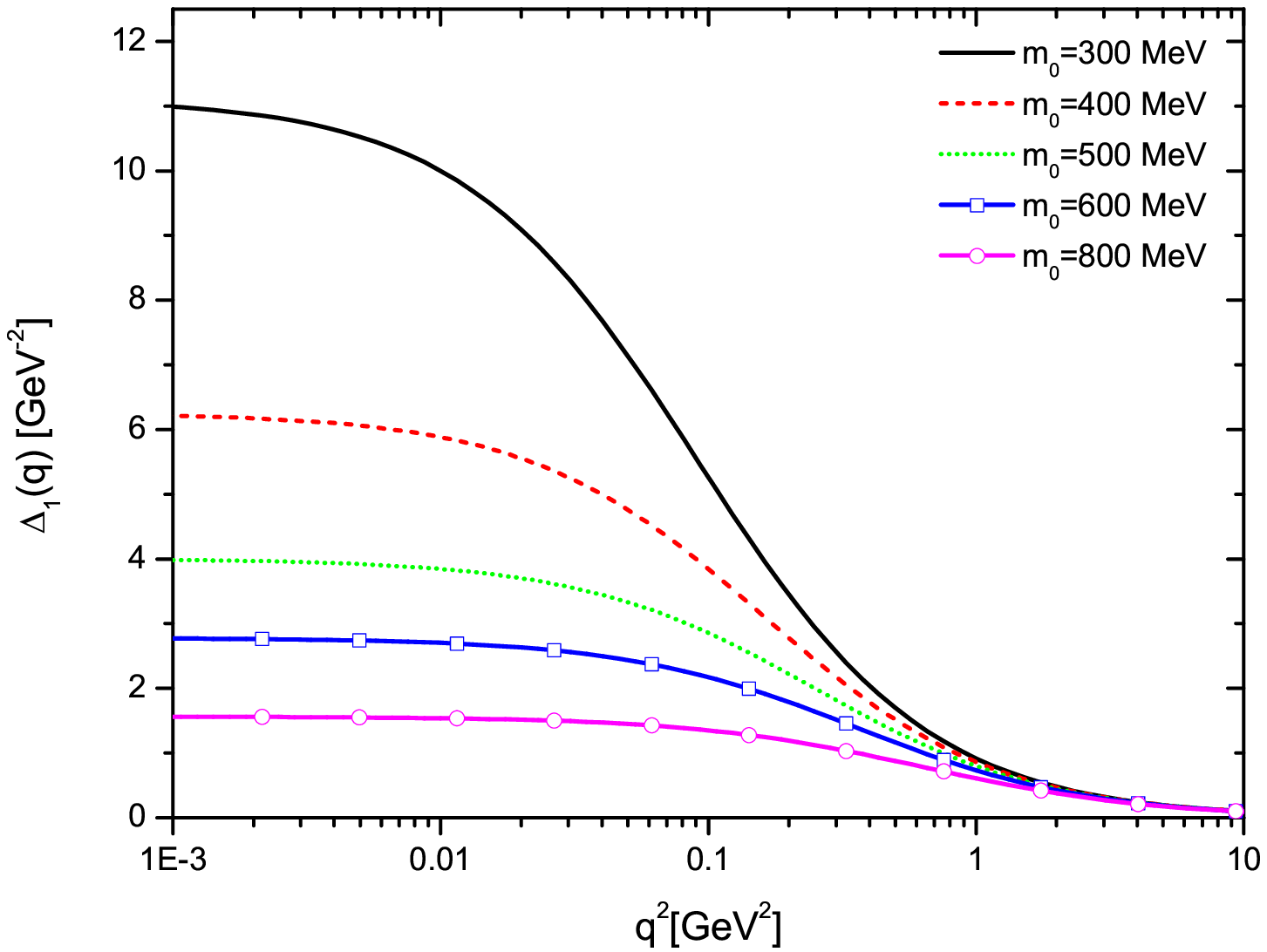}
\end{minipage}
\begin{minipage}[b]{0.50\linewidth}
\noindent
\includegraphics[scale=0.55]{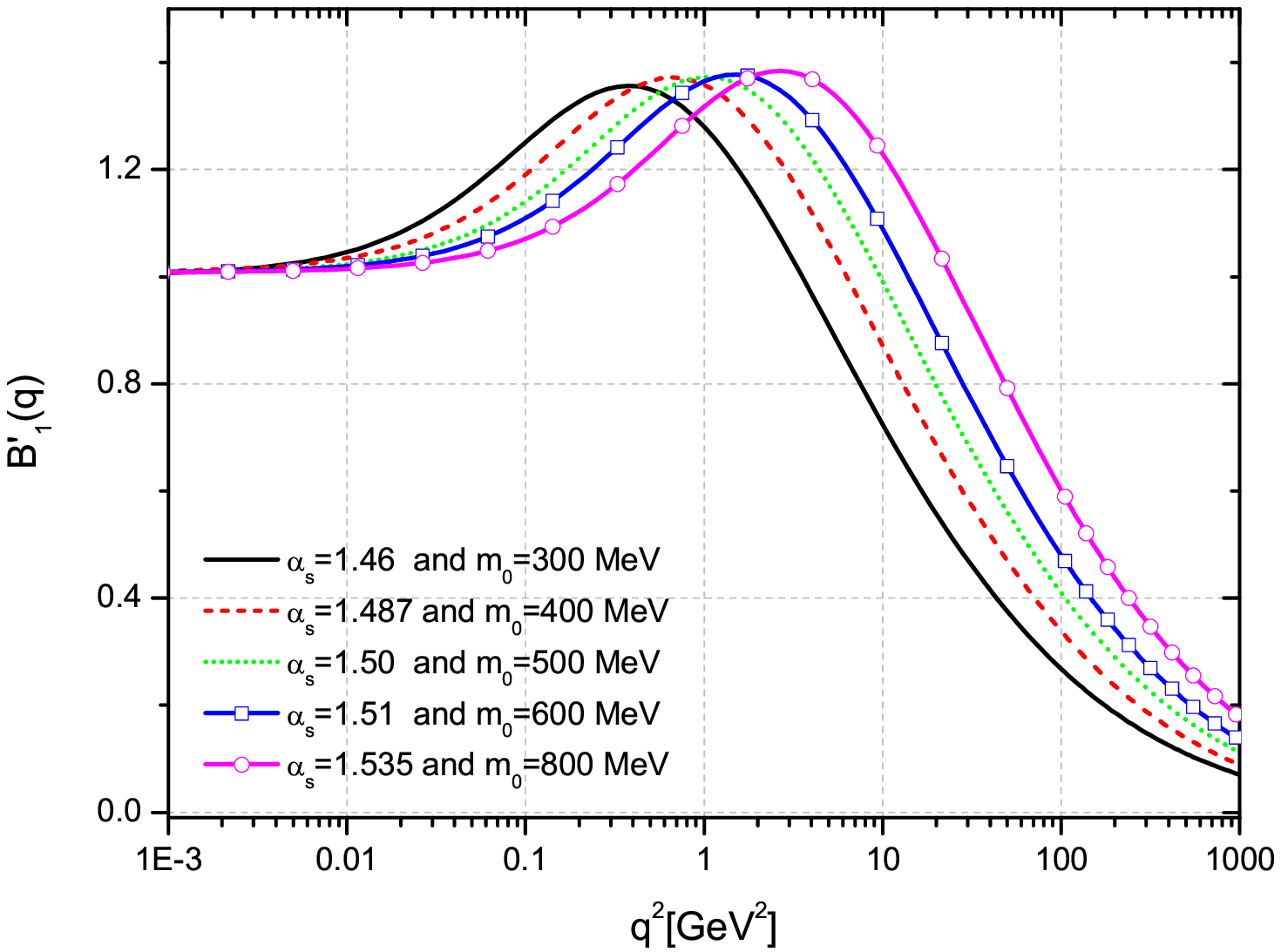}
\end{minipage}
\caption{{\it Left panel}: The behavior of the  gluon propagator $\Delta_1(q)$, given 
by  Eq.~(\ref{smassive}), for various values of 
$m_0$ in the range of $300-800 \, \mbox{MeV}$. 
{\it Right panel}: The corresponding solutions for $B'_1(q)$, 
obtained using the gluon propagators shown on the left panel. For each value of 
$m_0$, we found that the solution for $B'_1(q)$ is obtained for a particular value of $\alpha_s$.}
\label{vary_m}
\end{figure}

\begin{figure}[!t]
\begin{center}
\includegraphics[scale=0.55]{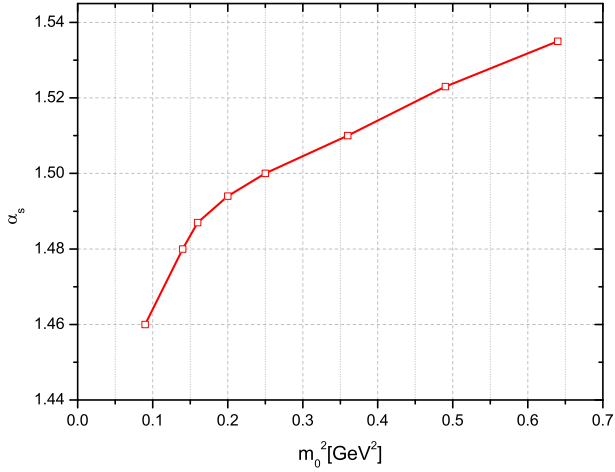}
\end{center}
\vspace{-1.0cm}
\caption{The values of $\alpha_s$ furnishing non-trivial solutions to Eq.~(\ref{weakBS})
as we vary $m_0$ in the $\Delta_1(q)$ of  Eq.~(\ref{smassive}).}
\label{a_m}
\end{figure}

{\it (v)} On the left panel of Fig.~\ref{vary_m} we plot $\Delta_1(q)$, given by Eq.~(\ref{smassive}),
for different values of  $m_0$ in the range of \mbox{$300-800 \, \mbox{MeV}$}. In order to determine  
how the solutions are modified when 
one varies the value of $m_0$, we show on the right panel of Fig.~\ref{vary_m} 
the various $B'_1(q)$, all of them 
normalized at $B'_1(0)=1$. As we can see, the solutions  display the same qualitative 
behavior; however, for each $m_0$, the non-trivial solution is obtained  
for a different value of $\alpha_s$.  
In fact, as the values of $m_0$ increase, so do the values of $\alpha_s$ 
needed for obtaining a solution; the exact  dependence of $\alpha_s$ on $m_0^2$ is shown in Fig.~\ref{a_m}.

\begin{figure}[!b]
\begin{minipage}[b]{0.45\linewidth}
\noindent
\centering
\hspace{-1cm}
\includegraphics[scale=0.55]{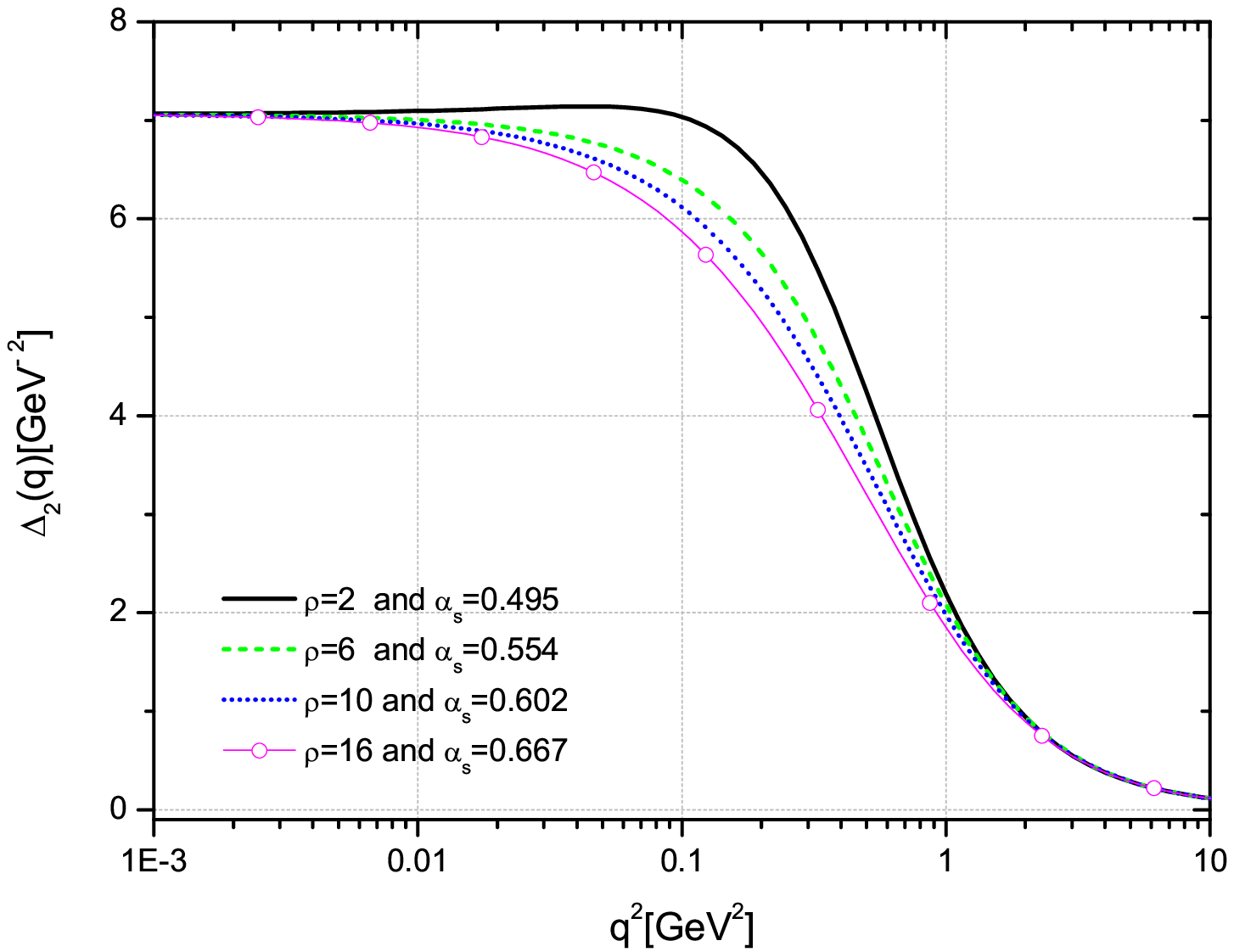}
\end{minipage}
\begin{minipage}[b]{0.50\linewidth}
\noindent
\includegraphics[scale=0.55]{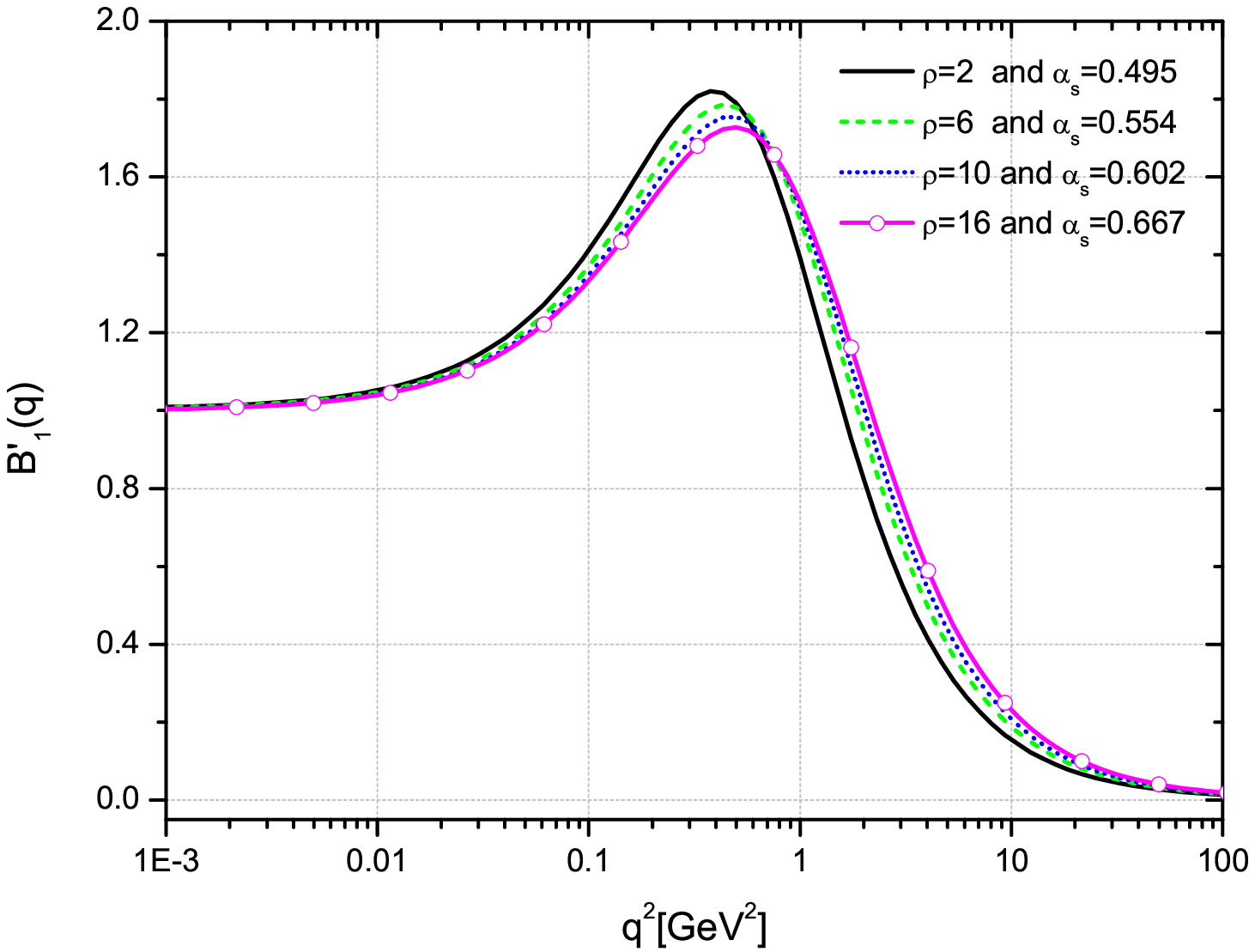}
\end{minipage}
\caption{{\it Left panel}: The behavior of the  gluon propagator, $\Delta_2(q)$, given 
by Eq.~(\ref{gluon2}), when the value of  $m_0=367\, \mbox{MeV}$ is fixed, and $\rho$
varies in the range 2-16.
{\it Right panel}: The corresponding solutions for $B'_1(q)$ 
obtained with the gluon propagators shown on the left panel. To each value of $\rho$ 
corresponds a specific value of $\alpha_s$ that yields a solution $B'_1(q)$.}
\label{vary_rho}
\end{figure}

\begin{figure}[!t]
\begin{center}
\includegraphics[scale=0.55]{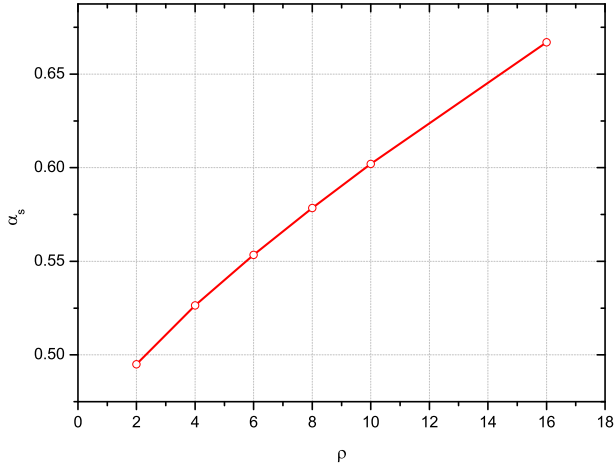}
\end{center}
\vspace{-1.0cm}
\caption{The values of $\alpha_s$ 
for which we obtain non-trivial solutions to Eq.~(\ref{weakBS})
as we vary $\rho$ in $\Delta_2(q)$ of  Eq.~(\ref{gluon2}).}
\label{a_rho}
\end{figure}

{\it (vi)} Next, we study how size variations in the intermediate region 
of the gluon propagator change the values of $\alpha_s$ needed in order to  
obtain non-trivial solutions from Eq.~(\ref{weakBS}). To 
address this point systematically, we employ the
gluon propagator $\Delta_2(q)$ of Eq.~(\ref{gluon2}), 
varying the  parameter $\rho$
in the range of $\rho \in [2,16]$, keeping fixed $m_0=367\, \mbox{MeV}$, 
as shown on the left panel of Fig.~\ref{vary_rho}; the corresponding 
$B'_1(q)$ for each value of $\rho$ are plotted on the right panel.
Evidently, decreasing $\rho$ increases the support of the gluon propagator 
in the intermediate region, and, as a result, one needs smaller value of $\alpha_s$ 
in order to obtain solutions for $B'_1(q)$.
This last property is better seen in Fig.~\ref{a_rho}, where we present 
the  values of $\alpha_s$  needed to solve Eq.~(\ref{weakBS})
as one varies $\rho$ in $\Delta_2(q)$.

\subsection{Non-linear treatment:~uniqueness of $B'_1(x)$ and $m^2(x)$}
\label{cc3}

In the previous subsection we have practically solved Eq.~(\ref{weakBS}) in isolation, 
in the sense that we have not used the supplementary conditions of Eq.~(\ref{exeq}), 
and have treated  $\Delta(q)$ as an external independent quantity.
As a result, the homogeneous Eq.~(\ref{weakBS}) was effectively linearized, 
giving rise to families of solutions $cB'_1(x)$, parametrized by the value of $c$.  
In this subsection we will restore the non-linearity of Eq.~(\ref{weakBS}); as a result,   
the arbitrariness in the value of $c$ is completely 
eliminated, and one obtains a single expression for $B'_1(x)$ and $m^2(x)$, 
for a unique value of $\alpha_s$.

\begin{figure}[!t]
\begin{center}
\includegraphics[scale=0.65]{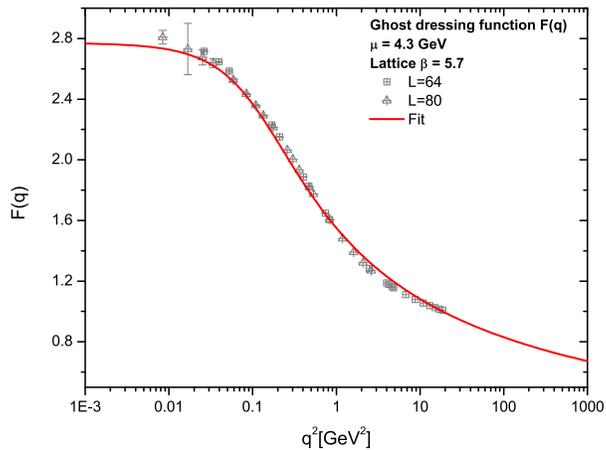}
\end{center}
\vspace{-1.0cm}
\caption {Lattice results~\cite{Bogolubsky:2007ud} 
for the ghost dressing function, $F(q)$, renormalized at \mbox{$\mu=4.3$ GeV}. Notice that
$F(0)\approx 2.76$.}
\label{lghost}
\end{figure}

The way a unique solution for $B'_1(x)$ is singled out [{\it i.e.}, a value for $c$
is dynamically chosen] is by combining  Eq.~(\ref{me-geneuc}) and Eq.~(\ref{i0_eu});
specifically, we will require that the value for ${\bar I}(0)$ obtained from the former equation 
coincides with that obtained from the latter, namely that 
\be
\sqrt{\Delta^{-1}(0)/{4\pi\alpha_s F^2(0)}} = \frac{3C_A}{32\pi^2} \int_0^{\infty} \!\!\!\! dy\, y^2\Delta^2(y)B'_1(y) \,,
\label{fixi}
\ee
Now, the lhs of (\ref{fixi}) is fixed, because, as mentioned in the previous subsection, we must have $\alpha_s=0.492$  
in order for Eq.~(\ref{weakBS}) to have solutions for this particular (lattice) 
propagator as input, while $\Delta^{-1}(0)$ and $F(0)$ are fixed 
from the lattice.  
Specifically, the  $SU(3)$ large-volume lattice 
simulations of  Ref.~\cite{Bogolubsky:2007ud} yield \mbox{$\Delta^{-1}\approx 0.141$} (see Fig.~\ref{props})
and $F(0)\approx 2.76$ (see Fig.~\ref{lghost}).
Then, the integral on the rhs (\ref{fixi}) must match the value of the lhs, and this can only happen for one 
particular member of the family $cB'_1(x)$.

\begin{figure}[ht]
\begin{center}
\includegraphics[scale=0.55]{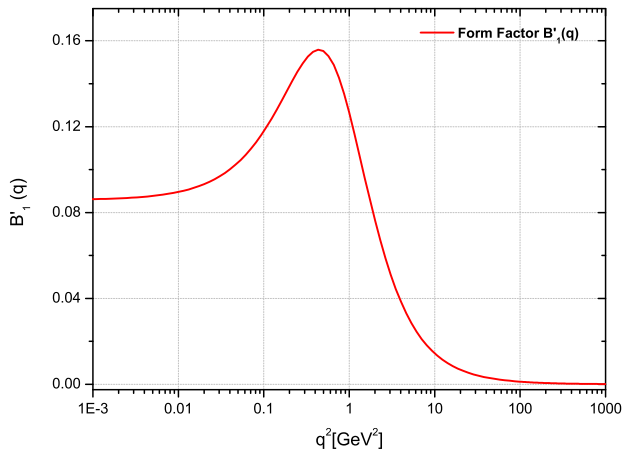}
\end{center}
\vspace{-1.0cm}
\caption{The numerical solution $B'_1(q)$ 
obtained from Eq.~(\ref{weakBS}), under the constraints imposed  
by Eqs.~(\ref{me-geneuc}) and~(\ref{zero}), and with  $\alpha_s=0.492$.}
\label{sol_int}
\end{figure}

In Fig.~\ref{sol_int}, we show the solution for $B'_1(x)$, which satisfies the constraints imposed on the 
value of $I(0)$,  obtained when $\alpha_s=0.492$ and $B'_1(0)= 0.086$.
 
\begin{figure}[!t]
\begin{center}
\includegraphics[scale=0.55]{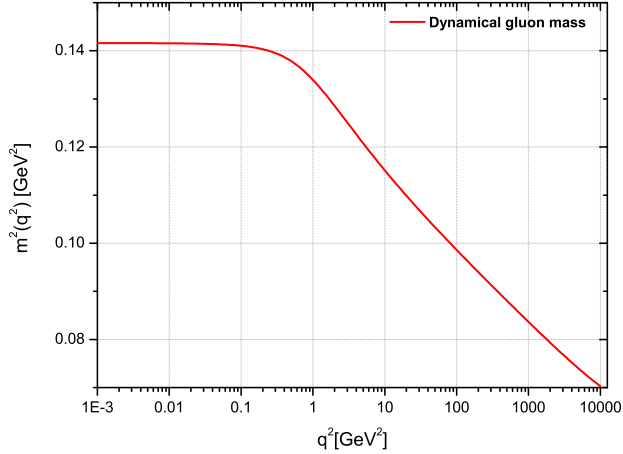}
\end{center}
\vspace{-1.0cm}
\caption{The square of the dynamical gluon mass,  
obtained from Eq.~(\ref{massrelation}), after plugging into it   
the $B'_1(x)$  shown in Fig.~\ref{sol_int}.}
\label{sol_mass}
\end{figure}

Once the unique solution $B'_1(q)$ has been determined, one may use  Eq.~(\ref{exeq}) to 
determine the behavior of the  squared gluon mass $m^2(x)$. 
Integrating numerically $B'_1(q)$  and fixing $m(0)=0.14$, we obtain the result shown in Fig.~\ref{sol_mass}.

Evidently, the function  
$m^2(q^2)$  displays a plateau in the deep infrared, 
and then  decreases sufficiently fast in the ultraviolet region, 
as expected on general grounds~\cite{Cornwall:1981zr,Aguilar:2009ke,Aguilar:2011ux}.

\newpage

\section{\label{dec}Decoupling of the massless excitation: an example}

In this section we give an explicit example of how the massless excitation 
decouple from an on-shell amplitude. Specifically, we will show 
how this is indeed what happens in the case of the four-gluon amplitude.
To be sure, a complete proof of the decoupling of the massless excitation  
from all Yang-Mills amplitudes 
requires the treatment of kernels 
with an arbitrary number of incoming gluons, which is beyond our powers at present. 
However, the example considered here captures the essence of the underlying 
decoupling mechanism.  

\begin{figure}[!b]
\center{\includegraphics[scale=0.5]{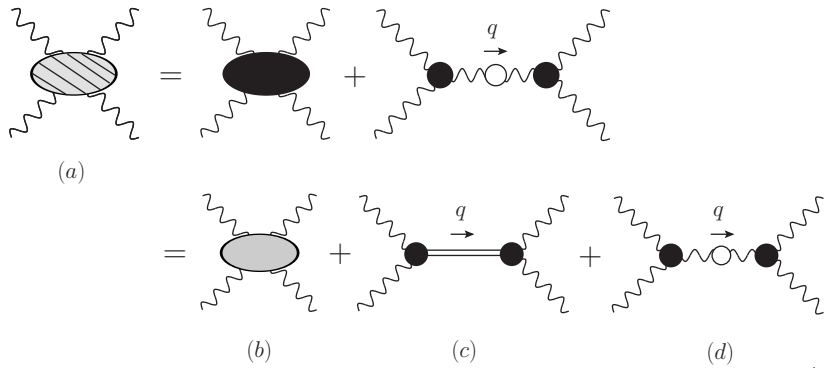}}
\caption{The complete four-gluon amplitude and the various terms composing it.}
\label{poledecoupling}
\end{figure}

The demonstration followed here is similar to that given in \cite{Jackiw:1973ha} 
for the case of an Abelian model.
One starts by considering the {\it complete} four-gluon amplitude, 
[graph $(a)$ in Fig.~\ref{poledecoupling}], which 
consists of three distinct pieces: (i) the amplitude represented by the diagram  $(b)$, 
which is regular as $q^2\to 0$, 
(ii) the graph $(c)$,  which contains the massless excitation, coupled to the external gluons through the 
proper vertex function $B$, and (iii) the one-particle reducible term, denoted by  $(d)$, which is   
excluded from the SDE kernel in the usual skeleton expansion. 
Of course, the above amplitudes are none other than 
$(b_2)$,  $(b_3)$, and $(a)$ in Fig.~\ref{Fullkernel}, respectively.      
  Since the amplitude $(b)$ is regular by construction, one must only demonstrate that, 
as $q^2\to 0$, the divergent part of $(c)$, whose origin is the massless excitation,   
cancels exactly against an analogous contribution contained in  $(d)$, 
leaving finally a regular result.

We start by considering the term $(d)$. Within the PT-BFM framework that we use, 
the off-shell gluon (carrying momentum $q$) is effectively converted into a 
background gluon; thus, 
the gluon propagator appearing inside ($d)$ is given by $\widehat\Delta(q)$, 
while the two three-gluon vertices are the $\NV$ defined in Eq.~(\ref{NV}). 
So, 
\bea 
(d) &=& -i g^2\, \NV_{\alpha\mu\nu}(q,p_1,p_2) P^{\alpha\beta}(q) \widehat\Delta(q) 
\NV_{\beta\rho\sigma}(q,p_3,p_4) 
\nonumber\\
 &=& -i g^2 \, \NV_{\alpha\mu\nu}(q,p_1,p_2) 
\widehat\Delta(q) {\NV}^{\alpha}_{\rho\sigma}(q,p_3,p_4) \,,
\label{dec1}
\eea 
where the factor $(-i)$ comes from the definition of the gluon propagator, Eq.~(\ref{prop}). 
In the second line we have eliminated the longitudinal term $q^{\alpha}q^{\beta}/q^2$   
inside $P^{\alpha\beta}(q)$ using the ``on-shellness'' condition 
\bea
q^{\alpha} \NV_{\alpha\mu\nu}(q,r,p) |_{{\rm o.s.}} &=&
[\Delta^{-1}({p^2})P_{\mu\nu}(p) - \Delta^{-1}({r^2})P_{\mu\nu}(r)]_{{\rm o.s.}}
\nonumber\\
 &=& 0 \,,
\label{dec2}
\eea
valid for both three-gluon vertices. 
We emphasize that 
the full  $\NV$ is needed (with the $V$ included) for the on-shellness condition 
of  Eq.~(\ref{dec2}) to be fulfilled. Note also that, 
if one had chosen  a non vanishing gauge-fixing parameter $\xi$  
for the gluon propagator (instead of the $\xi=0$ of the Landau gauge),  then  
the condition of Eq.~(\ref{dec2})
is instrumental for the cancellation of the unphysical parameter $\xi$ from the physical amplitude.

Next, it is clear that from the vertex $V$ contained in $\NV$ only the part $U$ survives, 
[see Eq.~(\ref{VRU})], because all longitudinal momenta contained in $R$ are annihilated 
on shell, i.e. when contracted with the appropriate polarization vectors $e_{\mu}(p)$, 
due to the validity of the relation $p^{\mu}e_{\mu}(p) =0$.  
Then,  we have that (suppressing indices)
\bea
\NV \widehat\Delta \NV &=&  (\bqq+U) \widehat\Delta (\bqq+U)
\nonumber\\
&=& \bqq \widehat\Delta \bqq + \NV \widehat\Delta U + U \widehat\Delta \NV - U \widehat\Delta U \,.
\label{dec3}
\eea
Given that the first term in (\ref{dec3}) is regular, while 
the second and third term vanish on shell by virtue of (\ref{dec2}) [which is triggered because 
$U_{\alpha\mu\nu}$ is proportional to $q_{\alpha}$, see  Eq.~(\ref{VU})], 
we are led to the following expression for the pole part of $(d)$  
\be
(d)_{\rm pole} = i g^2 U_{\alpha\mu\nu} \widehat\Delta(q) U^{\alpha}_{\rho\sigma}\,.
\label{dec4}
\ee
Then, using Eqs.~(\ref{VwB}) and (\ref{qI}), we obtain 
\be
(d)_{\rm pole} =  - \left\{B \left(\frac{i}{q^2}\right) B\right\} [g^2 I^2(q)\widehat\Delta(q)] \,.
\label{dec5}
\ee
Now, in the limit $q^2\to 0$, the quantity in square brackets goes to $1$, 
precisely by virtue of Eq.~(\ref{PTresfin}) [remember, $\widehat\Delta^{-1}(0) = {\widehat m}^2(0)$]. 
Therefore, 
\be
\lim_{q^2\to 0} (d)_{\rm pole} = - \lim_{q^2\to 0} \left\{B \left(\frac{i}{q^2}\right) B \right\} \,,
\label{dec6}
\ee
which is precisely the contribution of the term $(c)$ in the same kinematic limit, 
but with the opposite sign. Therefore, the on-shell four-gluon amplitude
is free from poles at $q^2 = 0$, as announced. 

Finally, note that, as an alternative, one might opt 
for eliminating completely any reference to $V$ in the amplitude $(d)$ 
from the very beginning, namely 
the first step in  Eq.~(\ref{dec1}); this is of course possible, 
given that some parts of 
the fully longitudinal vertex $V$
vanish on shell, while the rest vanishes when contracted with the transverse projector 
$P_{\alpha\beta}(q)$.
In such a case, however, one may not
dispose of the longitudinal part of $P_{\alpha\beta}(q)$ any longer, because now the 
on-shellness condition of Eq.~(\ref{dec2}) is distorted, 
precisely due to the absence of $V$. 
It is a straightforward exercise to 
demonstrate that if one were to take the produced mismatch into account, 
one would recover exactly the same result found above.

\section{\label{conc}Discussion and Conclusions}

The gauge-invariant generation of a gluon mass 
relies on the existence of massless bound-state excitations,  
which trigger the Schwinger mechanism. The presence of these 
excitations in the skeleton expansion of the full  
three-gluon vertex $\NV_{\alpha\mu\nu}$
induces longitudinally coupled pole structures,  
giving rise to a purely nonperturbative component, 
the pole vertex $V_{\alpha\mu\nu}$.   

In this article we have studied 
in detail the dynamical ingredients associated with the vertex $V$; 
in particular, the poles in $V$ are identified with the  
propagator of the massless scalar excitation, while the 
tensorial structure is determined by two basic purely nonperturbative quantities: 
the  transition amplitude, denoted by $I_{\alpha}$, which at the diagrammatic level 
 connects the gluon propagator with the 
massless scalar propagator, 
and the effective vertex, denoted by $B_{\mu\nu}$, 
 connecting the 
 massless excitation to two outgoing gluons. 

The powerful requirement of maintaining the gauge invariance of the theory intact 
restricts the form of the various form factors composing $B_{\mu\nu}$, and establishes 
a non-trivial connection between the  transition function and the first derivative 
of the momentum-dependent gluon mass. 
The insertion of the vertex  $V_{\alpha\mu\nu}$ (or, effectively, its 
surviving component $U_{\alpha\mu\nu}$)
 into the SDE of the gluon propagator (in the Landau gauge) allows one to express, 
at zero momentum transfer, the gluon mass in terms of the transition function, 
by means of a rather simple formula. In fact, it turns out that, 
the relevant dynamical quantity is the 
derivative of the form factor $B_1$, denoted by $B_1'$.

As we have demonstrated, in the 
aforementioned kinematic limit, 
the homogeneous BSE obeyed by the  $B_{\mu\nu}$ reduces in a natural way to 
an analogous integral equation for $B_1'$. 
The detailed numerical study of an approximate version of this latter equation  
reveals the existence of non-trivial solutions for $B_1'$, 
which, when inserted into the corresponding formulas, furnish  
the momentum dependence of the gluon mass. The existence of these solutions 
adds weight to the hypothesis 
that the nonperturbative Yang-Mills dynamics lead indeed to the 
formation of the required massless bound-states.

It is clear that some of the dynamical aspects of this problem merit a
further detailed study, due to their relevance in the ongoing scrutiny
of  the  infrared  properties  of the  Yang-Mills  Green's  functions.
Particularly important  is to consider the effects of  
bound-state poles in the SD kernels of not only the three-gluon vertex, 
as we did in this article, but of all other 
fundamental  vertices  of  the  theory. 
Such  an  investigation  would
involve some  or all of the vertices  appearing in Eq.~(\ref{restU}),
which would  form a coupled system of  homogeneous integral equations.
Given the
recent lattice  results on  the ghost propagator,  
especially interesting  in this  context is the  dynamical information
that  one  might be  able  to obtain  about  the  quantity $B$,
corresponding to the wave-function of the ghost-ghost channel 
[(vertex $a_1$ in Fig.~(\ref{Uexpansion})]. Specifically, while
the ghost dressing  function $F$ is found to be finite  in the infrared, 
a fact that  can be  explained by  the presence of  the gluon  mass that
saturates  the  corresponding  perturbative  logarithm,  there  is  no
dynamical  mass associated with  the ghost  field.  One  would expect,
therefore, that  the solution of the corresponding  system should give
rise to a non-vanishing $B_1'$, as before, but to a vanishing 
ghost-ghost wave function $B$.

As has been explained in detail in Section \ref{gc}, the incorporation 
the massless excitations modify the three-gluon vertex $\bqq$, giving rise to the 
new vertex $\NV$, defined in  Eq.~(\ref{NV}).
It would certainly be particularly interesting to compare the 
characteristic features of $\NV$ 
 with results obtained on the lattice for the three-gluon vertex~\cite{Cucchieri:2008qm}.
In particular, one might in principle be able to relate the presence  
of the massless poles to possible   
divergences of some of the form factors, in the appropriate kinematic limit.
To that end, one must first determine the closed expression for the entire vertex 
$V$ from Eq.~(\ref{totlon}) and the WI and STIs it satisfies.
Then the answer should be written  in a standard basis, such that of Ball and Chiu~\cite{Ball:1980ax,Binosi:2011wi}, 
and  the final result projected on the particular kinematic configurations 
usually employed in the lattice calculations. 
We hope to be able to carry out this project in the near future.

\acknowledgments 

The research of D.I. and J.P. is supported by the European FEDER and  Spanish MICINN under 
grant FPA2008-02878, and that of  V.M. under grant FPA2010-21750-C02-01.
The work of  A.C.A.  is supported by the Brazilian
Funding Agency CNPq under the grant 305850/2009-1 and project 474826/2010-4.

\end{document}